**Integrating Large Language Models with Human Data to Design Personalized Decoy Nudges**

Vladimir Maksimenko[1], Leonard Lee[2*], and Prateek Bansal[1*]

[1]Department of Civil and Environmental Engineering, National University of Singapore, Singapore 117576
[2]NUS Business School, Lloyd's Register Foundation Institute for the Public Understanding of Risk, National University of Singapore, Singapore 119245

*Corresponding Authors: leonard.lee@nus.edu.sg; prateekb@nus.edu.sg

***Acknowledgment***
The authors thank Dr. Prateek Gupta from the Max Planck Institute for Human Development, Dr. Qingyao Xin from the National University of Singapore and Prof. Bin Zhang from the School of Management at Beijing Institute of Technology for helpful discussions that contributed to the development of the initial research concept. Parts of this work were presented at Theory and Practice in Marketing (TPM-2026) conference.

***Declaration of conflicting interest***
The authors declare that they have no known competing financial interests or personal relationships that could have appeared to influence the work reported in this paper.

***Funding statement***
This research is supported by the Presidential Young Professorship grant to Prateek Bansal

***Ethical approval and informed consent statements***
The ethical aspects of this research have been approved by the Institutional Review Board of the National University of Singapore (Reference Code: NUS-IRB-2024-774). Respondents provided informed consent to participate in the study.

***Data availability statement***
Anonymized human choice data, LLM-generated responses, experimental materials, and analysis code will be publicly shared upon acceptance of the article. They can also be made available to reviewers during the review process, if needed.

***Open practices statement***
The human study has been preregistered (OSF: https://osf.io/u8kyv). Experiment design, hypotheses, and data collection plan are specified prior to data collection.

***Use of generative AI***
OpenAI's GPT-4o-mini was used to generate simulated choice responses, as described in the Methods section. ChatGPT (OpenAI) was also used to assist with manuscript grammar editing. The authors reviewed and verified all AI-assisted content and take full responsibility for the research and the final manuscript.

## Abstract

Although nudge strategies can influence behavior, their effectiveness varies across individuals, highlighting the need for personalized nudge design. We hypothesize that large language models (LLMs) can facilitate the development of individual-specific nudges while reducing the financial and time costs associated with human data collection by augmenting small human samples or by replacing human participants. Using the decoy effect as a test case, we evaluate these hypotheses in a preregistered field experiment involving 3,495 participants across five countries. We show that decoy parameters identified by LLMs as optimal for a given country generate stronger nudging effects than non-optimal parameters in the United States and, marginally, in Germany, but not in China, India, or Singapore. We find no evidence that LLM-informed personalized decoys produce a stronger nudging effect than country-level optimal decoys. Finally, we demonstrate that LLM-generated data can serve as informative priors for Bayesian choice models, improving predictive performance and parameter estimation with smaller human samples. At matched predictive performance, this approach yields estimated reductions of 37–75% in human sample size. Overall, our findings suggest that LLMs hold considerable promise as low-cost tools for piloting and optimizing personalized nudge strategies, particularly when complemented with a small sample of human participants.

*Keywords*: Decoy effect, Large Language Model, Choice models, Nudges

## Introduction

Nudges guide individual behaviors by subtly altering the decision-making environment. By modifying contextual cues, they draw attention to a desired option while leaving the characteristics of the available alternatives unchanged and preserving freedom of choice. Evidence from diverse domains, including marketing, healthcare, food choice and sustainability, demonstrates that nudges can strategically influence decision-making (Dai et al. 2021; Ghose et al. 2024; Huang et al. 2024; Mrkva et al. 2021; Mrkva et al. 2025; Romero and Biswas 2016).

Although nudge strategies are promising tools to achieve societal targets, their effectiveness is constrained by heterogeneity in consumer behaviors (Huber et al., 2014). Consequently, strategies that work well for some individuals may fail or even be counterproductive for others (Sunstein, 2022; Zimmermann et al., 2024). This variability diminishes the overall effectiveness of nudges if they are applied uniformly at the population level. Designing effective interventions therefore requires identifying *personalized* nudge parameters by *predicting* whether a given strategy will succeed for the target consumer segment (Mousavi et al., 2023; Szaszi et al., 2025). Predicting context-induced irrational behavior requires the estimation of behavioral models (Dumbalska et al. 2020; Landry and Webb 2021; Noguchi and Stewart 2018; Rooderkerk et al. 2011; Soltani et al. 2012) using data collected from different consumer segments. Such data collection requires substantial time and financial resources, thereby diminishing the economic efficiency of nudge strategies for behavioral policy design.

Large Language Models (LLMs) offer a potential means of addressing this challenge (Ye et al. 2025). Trained on vast amounts of human-generated text, LLMs have been shown to mimic aspects of human reasoning (Acerbi et al. 2023; Hagendorff et al. 2023; Webb et al. 2023) and predict responses in experiments across psychological and economic domains (Binz et al. 2025; Brand et al. 2023; Cui et al. 2025; Dillion et al. 2023). Even in tasks where LLMs fail to accurately replicate the behavior of real consumers (Goli and Singh, 2024; Peng et al. 2026), there may still be a potential for LLM–human alignment (Liu et al. 2024; 2025) using training sample of human data. This alignment could reduce the cost of experimentation by lowering sample size requirements, alleviating the need to collect representative data from all consumer segments. Finally, LLMs can incorporate individual profiles to generate tailored behavioral interventions. Even simple forms of personalization (based on gender, age, ethnicity, education, employment status or political affiliation) enable LLMs to produce persuasive messages that resonate with specific backgrounds and demographics (Argyle et al. 2025; Salvi et al. 2025). This emerging evidence implies that LLMs could play a valuable role in early-stage nudge design by acting as partial proxies for human respondents with heterogeneous preferences, thereby reducing reliance on large-scale human experiments.

In this study, we investigate whether LLMs can be used to assess the effectiveness of decoy-based nudging strategy and inform personalized decoy parameters. The decoy effect is a well-established concept in consumer psychology, whereby introducing a less attractive option can shift preferences toward a particular original option (Adler et al. 2024; Devine et al. 2025; Dumbalska et al. 2020; Evangelidis et al. 2024; Kim et al. 2025; Padamwar and Dawra 2024; Polman and Maglio 2024; Spektor et al. 2021; Wedell et al. 2022). A classic example is Starbucks' pricing strategy for coffee sizes: to make the large size (the target) more appealing relative to the small size, Starbucks introduced a medium size (the decoy) that was priced only slightly lower than the large but offered substantially less volume. Such decoy-based strategies could be readily incorporated in contexts where consumers choose between multiple alternatives, such as digital booking platforms or online shopping websites that present several options side by side.

Considering aviation as a case in point, we show that an LLM (OpenAI's GPT-4o-mini) can be used to inform personalized decoy parameters that nudge air travelers to offset their flight emissions. In a preregistered experiment conducted with 3,495 real-world air travelers across five countries, we show that decoy parameters identified by LLMs as optimal for a given country generate stronger nudging effects than non-optimal parameters in the United States and, marginally, in Germany. By contrast, LLM-informed decoy parameters have no effect on travelers in China, India, or Singapore, highlighting a Western bias in the GPT model. We further show that this misalignment between LLM and human behavior can be substantially reduced by using LLM responses as informative priors in the Bayesian estimation of choice models. Specifically, we demonstrate that LLM-informed priors enable stable parameter estimation and improve predictive performance across all countries, even when only 2% of human training data are available.

These findings underscore the potential of LLMs to support the design of personalized nudges while highlighting the limitations of relying on LLMs as a full substitute for human respondents. Instead, our results suggest that the greatest value of LLMs may lie in hybrid human–AI frameworks, where LLM-generated knowledge is used to inform behavioral models and substantially reduce human data requirements.

## Theoretical Foundation

### *Notion of Optimal Decoy and Need for Personalization*

Despite the long history of research on the decoy effect (Evangelidis et al. 2024), its effectiveness and replicability across different contexts remain subjects of debate (Simonson, 2014). One frequently discussed issue concerns the extent to which the effect depends on the specific parameters of the decoy. Early work by Huber et al (2014) showed that not all decoy configurations are equally effective, highlighting the need for models that can predict the optimal decoy parameters that maximize the nudging effect.

Some studies have systematically explored ranges of decoy parameters and developed models to explain how these parameters shape the resulting nudging effect. They show that the strength of the decoy effect depends on the distance between the decoy and the target in the attribute parameter space. Most studies report that decoy effect strength increases with distance (Dumbalska et al. 2020; Kaptein et al. 2016; Landry and Webb 2021; Soltani et al. 2012). At the same time, some studies also report that when the distance between the decoy and the target is too large, decoys lose their effectiveness (Kaptein et al. 2016; Landry and Webb 2021).

Importantly, reported decoy effects are not uniform across individuals. Both Soltani et al. (2012) and Dumbalska et al. (2020) observe variation in the magnitude and direction of individual-level responses to the same decoys. These findings suggest that a decoy configuration identified as optimal at the aggregate level may not be equally effective for every individual. However, existing studies do not examine whether this variability can be systematically explained by demographic or cultural characteristics, leaving an important gap in segment-specific decoy design.

### *LLM as a Human Participant*

In recent years, LLMs have demonstrated the promising potential to generate responses similar to those of human participants across a variety of experimental settings (Dillion et al. 2023), opening new opportunities for marketing research (Brand et al. 2023; Sarstedt et al. 2024). For instance, LLMs have been shown to replicate experiments from psychology, economics, and management science, achieving 81% replication rate for main effects and 63% replication for interaction effects (Cui et al. 2025). Similarly, Toubia et al. (2025) report that the accuracy of LLM predictions is 71.72% for experiments from the marketing and behavioral economics literature.

On the other hand, recent work in marketing and consumer research has also reported failures of LLMs to reproduce behavioral patterns of humans. For instance, Goli and Singh (2024) use LLMs to measure intertemporal preferences and report that LLMs exhibit larger discount rates than those observed among human subjects. Finally, a recent work by Peng et al. (2026) report that the correlation between LLM and human responses is weak potentially due to hyper-rationality, which limits the ability of LLMs to replicate irrational behavioral phenomena such as attraction and compromise decoy effects.

These findings suggest that LLMs can partially replace human participants for hypothesis testing in marketing and psychological science. While concerns remain about full replacement (Cummins 2025; Schröder et al. 2025), scholars broadly agree that LLMs can serve as an alternative to human participants in pilot experiments, with subsequent validation using real respondents.

Finally, the usefulness of LLMs for designing and evaluating decoy nudges remains largely unexplored. While the recent study of Peng et al. (2026) found no evidence that LLMs exhibit the decoy effect, their analysis was limited to a specific set of decoy parameters, and the generalizability of this finding remains unclear. It is therefore still an open question whether LLMs are susceptible to decoys, and whether they can be used to identify optimal decoy parameters for specific respondent segments to maximize nudging effectiveness.

### *LLMs – Human Behavioral Alignment*

The existing literature provides conflicting evidence regarding the ability of LLMs to accurately simulate human behavior. Consequently, several methods have recently been proposed to improve the alignment between LLM-generated and human responses.

The first strategy, few-shot learning, leverages an LLM's ability to learn from a small set of examples provided in the prompt (Brown et al. 2020; Chamieh et al. 2024; Song et al. 2023). The model extracts patterns from these examples and applies them to new situations to simulate human behavior (Kim et al. 2024). The second strategy is fine-tuning, which involves training a pre-trained LLM on a dataset of observed human behaviors and their corresponding contexts (Brand et al. 2023). During this process, the model parameters are updated so that its outputs better align with human behavior (Binz et al. 2025). However, fine-tuning is computationally demanding, may require substantial hardware resources, and does not necessarily provide an advantage over standard GPT models (Peng et al. 2026). The third strategy is based on the idea that prompts should be enriched with personas—representations of individuals' preferences, behavioral traits, and values that are not explicitly captured by standard socio-demographic variables (Park et al. 2023). These personas are usually inferred from a dataset of observed human choices and then incorporated into prompts when the LLM is asked to predict future decisions (Liu et al. 2024; 2025). A fourth strategy is statistical data augmentation (Wang et al. 2026). Using a subset of choice tasks answered by both humans and LLMs, a model learns to predict human choice probabilities from the LLM responses and the task attributes. This mapping is then applied to LLM responses for tasks where human data are unavailable.

These results suggest that alignment between LLM-generated and human responses can be improved using some amount of human data. At the same time, research in economics and marketing is often concerned not only with predicting final responses but also with estimating behavioral models whose parameters have clear economic interpretations and reflect latent preferences (Wang et al. 2026). In this context, a promising application of LLMs is the use of their responses as informative priors in Bayesian estimation of behavioral models when human data are scarce (Riegler et al. 2025). While this approach has shown promise in several domains (Arai et al. 2025; Evans et al. 2025; Sekulovski et al. 2026), it has not yet been applied to the estimation of choice models. Therefore, it remains an open question whether LLM-derived priors can improve the estimation of choice model parameters, particularly for models designed to capture the decoy effect.

## Current Study

Drawing on the identified gaps, this study answers three research questions. First, we test whether LLMs exhibit the decoy effect and whether the strength of this effect varies across different decoy parameters. Second, we use LLMs to identify decoy parameters that maximize the nudging effect for different consumer segments and validate these recommendations in a human experiment. Third, we investigate whether the Bayesian framework can help align LLM-generated responses with observed human behavior with a small sample of human participants.

To examine whether LLM-based predictions align with human behavior, we designed a field experiment with air travelers in which decoy parameters were tailored to participants' country of residence and personal profiles according to the LLM's predictions. The experimental design was preregistered prior to the start of data collection. (OSF: https://osf.io/u8kyv/overview?view_only=6e140b187ec645e3be79215ae8e8f350).

To test the alignment framework, we used a choice model grounded in the marketing literature that captured preference shifts under the decoy effect (Rooderkerk et al. 2011) and extended it to account for heterogeneity in decoy susceptibility across consumer segments defined by demographic characteristics and attitudes. We evaluated the framework by comparing both predictive performance and parameter recovery under uninformative and LLM-informed priors across different levels of human data availability.

### *Study Context*

We employed a flight-booking scenario as a case study and examined whether decoy-based interventions could encourage voluntary $CO_2$ offset purchases (see Fig. 1).

Figure 1: Illustration of Choice Situations during a Hypothetical Flight Booking Scenario.

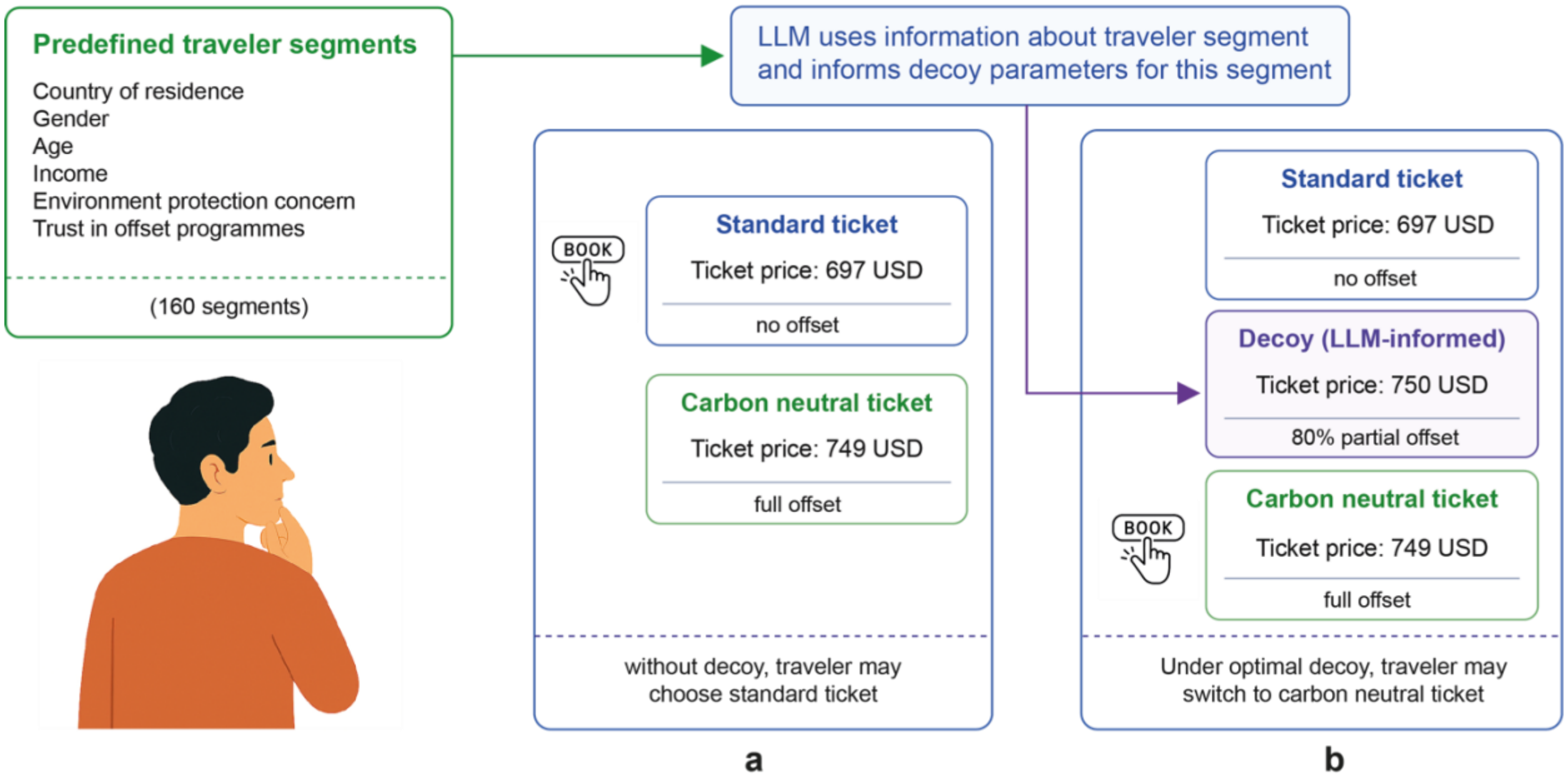

*Notes*: The task involves choosing between a standard ticket and a slightly more expensive carbon-neutral ticket (a). We analyze whether adding a third option (decoy) to the choice set (b) nudges air travelers to book the carbon-neutral ticket more frequently. The parameters of the decoy option are set by an LLM for each traveler based on their country of residence, demographic characteristics, and attitudes.

In the existing flight booking systems that present participants with a choice between a standard ticket offering no offset and a slightly more expensive carbon-neutral ticket offering full offset (the target), we propose introducing a third option (the decoy) that partially offsets emissions but at a higher relative price than carbon-neutral ticket. It is feasible to integrate such an intervention into existing flight booking systems as airlines provide multiple fare bundles with varying offset levels and in-flight services. Moreover, this information is usually presented in the form of a table, and the attribute values are shown as numbers (e.g., price and percentage of emission offset), which is beneficial for producing the decoy effect (Frederick et al. 2014; Yang and Lynn 2014)

We defined segments of travelers based on a brief personal profile that included country of residence, age, gender, income, and attitudes toward environmental protection and carbon offsetting programs. For simplicity, continuous variables such as age and income were converted into binary indicators using the country-specific median values. As a result, each segment was defined by the combination of attributes reported in Appendix A. Such definition resulted in 160 unique segments, 32 segments per country. We focused on the attraction decoy and identified the decoy parameters that maximized nudging effectiveness. For this purpose, we explored a parameter space defined by decoy price and offset, resulting in 45 distinct decoy configurations reported in Appendix B.

## Methods

This study involved three sequential steps: LLM-based simulations, field experiment with travelers, and LLM-human alignment as shown in Fig. 2.

In the LLM-based simulations, we used the predefined information about traveler segments to generate LLM responses across all decoy parameter combinations and choice situations. These responses were subsequently used to identify the decoy parameters that maximized the probability of carbon-offset purchases at both the country and traveler-segment levels.

In the field experiment, we collected approximately 3,600 responses from participants across five countries to the decoy configurations identified through LLM-based simulations. We then tested whether the decoys predicted by the LLMs to be optimal indeed maximized the decoy effect in human field data.

Finally, for the LLM–human alignment analysis, we used LLM responses to construct informative priors, which were incorporated into the Bayesian estimation of the choice model. By varying the proportion of human observations used for model estimation, we evaluated the extent to which LLM-informed priors could reduce the amount of human data required for accurate model estimation.

Figure 2: Flow Chart of The Study.

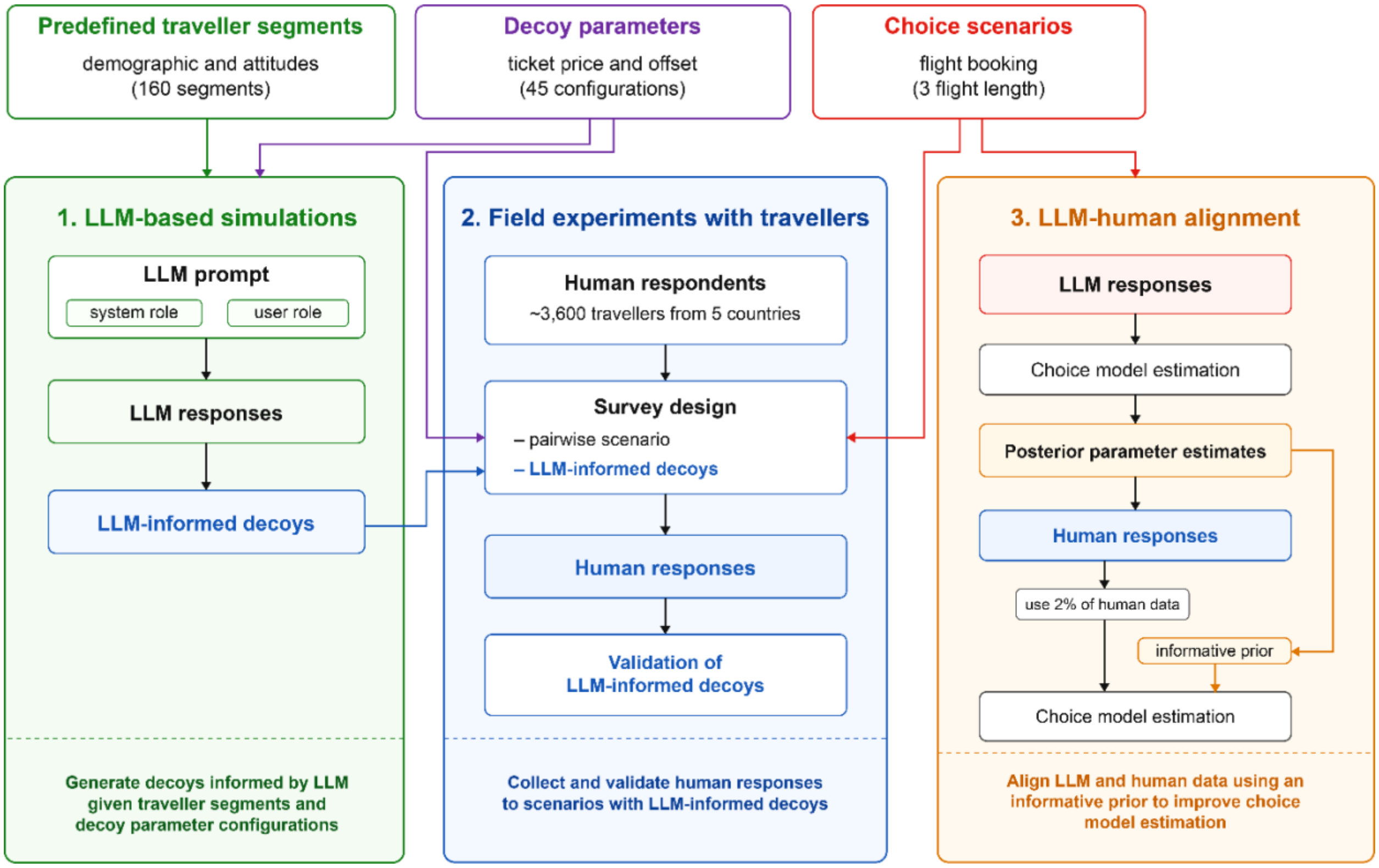

*Notes:* In the *LLM-based simulations*, responses were generated for all combinations of decoy parameters and traveler segments to identify optimal decoys. In the *field experiment with travelers*, human responses to these decoys were analyzed. In. *LLM–human alignment*, LLM responses were used to construct informative priors for Bayesian estimation of a choice model, which was then fitted using a small subset of human responses.

### *LLM-based Simulations*

We conducted experiments with LLMs using the OpenAI API and the gpt-4o-mini model to identify optimal decoy parameters for different air traveler segments. We used the model without fine-tuning and set the temperature to 0.8, while keeping all other parameters at their default values. To enable the model to account for the individual's personal profile, we specified a system role prompt (see Appendix C). Thus, 160 models were specified to represent all traveler segments. We provided each model with a flight booking scenario asking to choose one of the presented options: target, competitor, and decoy setting the user role prompt (see Appendix C).

In the LLM experiment, we first generated 30 flight booking scenarios with unique combinations of target and competitor prices using the algorithm described in Appendix D. For each scenario, we consider one pairwise choice situation involving only the target and the competitor, as well as 45 choice situations including a decoy, in which the target and competitor parameters remain unchanged, and the decoy parameters (*decoy_price* and *decoy_offset*) are set according to the algorithm described in Appendix B

Since LLMs are sensitive to the sequence of text (Brucks and Toubia, 2024) and prompt wording (Blanchard et al. 2025), we presented each choice situation four times with randomized ordering of the choice options. Because we expected LLM responses to the same question to vary due to high temperature, each situation was presented 25 times, resulting in 100 responses (4 situations with randomized order x 25 runs) per choice situation. Aggregating the choice situations across the 30 flight booking scenarios yields 3,000 responses per model for each of the 46 choice situations (one pairwise and 45 decoy conditions).

For the pairwise choice situation, we calculated the probability of choosing the target (offsetting probability) as the number of instances in which the LLM chose the target divided by 3,000 choice instances. In the decoy choice situations, we focused on the relative share of the target as a measure of decoy effectiveness. Accordingly, for each of the 45 decoys, the offsetting probability was calculated as the number of instances in which the LLM chose the target divided by the number of instances in which the LLM chose either the target or the competitor, excluding instances in which the decoy was chosen.

After performing these calculations for all 160 traveler segments, we obtained segment-level offsetting probabilities for the pairwise situation and for each of the 45 decoy situations. To quantify the preference shift for each segment, we subtracted the offsetting probability in the pairwise situation from the offsetting probability calculated for each decoy situation. For each segment, we selected the five decoys that exhibited the largest preference shifts; we refer to them as *LLM-informed individual-optimal decoys*. Next, we averaged the preference shifts for each decoy across all segments within each country to obtain country-level preference shift estimates. Based on these estimates, we selected the five decoys with the highest and the five with the lowest country-level preference shifts, which we refer to as *LLM-informed country-optimal* and *LLM-informed country-non-optimal decoys*, respectively. Finally, to demonstrate that the optimal decoy parameters vary across segments, we calculated, for each of the 45 decoys, the number of segments in which it was selected as one of the five optimal configurations.

***Field Experiment with Travelers***

To test the effectiveness of LLM-informed individual-optimal and country-optimal decoys, we recruited participants through professional survey panels. Eligible participants were at least 18 years old, had flown at least once in the past 12 months, and were born in and have continuously resided in Singapore, the United States, India, Germany, or China. Participants must also provide unambiguous binary responses for gender (male or female), age group (below or above the median), and income group (below or above the median). We targeted at least 600 valid respondents per country ($N \geq 3{,}000$ total), with balanced quotas for gender, age, and income within each country (for more details regarding power analysis and sample size estimation, see OSF preregistration:
https://osf.io/u8kyv/overview?view_only=6e140b187ec645e3be79215ae8e8f350). The study was approved by the Institutional Review Board of the authors' home institution, and all participants gave informed consent. Respondents were remunerated by the survey panel provider in accordance with its standard compensation procedures.

This field experiment was conducted via Qualtrics. Participants first provided informed consent and were then tested on their understanding of the concept of emission offsetting using a choice situation with a correct answer; failure to answer correctly led to exclusion. Participants who passed this check then completed screening questions (flight history and country of residence) and demographic questions (gender, age group, and income group), with quotas enforced in real time. They also answered two binary attitudinal questions (trust in offsetting programs and concern for environmental protection), which were used to assign participants to predefined segments according to Appendix A and to determine LLM-informed individual-optimal decoy parameters.

Participants who passed all screening and sanity checks were presented with three hypothetical flight booking scenarios (short-, medium-, and long-haul flights), with fully randomized ordering per respondent. Within each scenario, participants responded to five choice situations: pairwise choice, three decoy situations, and attention check. Thus, each participant encountered 15 choice situations in total. Across these situations, prices for the competitor and the target were set using the algorithm described in Appendix D. and remained unchanged. Decoy situations involved (i) Country-optimal decoy, with values inferred by the LLM to maximize offsetting probability at the country level; (ii) Country-non-optimal decoy, with values inferred by the LLM to reduce offsetting probability in that country; and (iii) Individual-optimal decoy, with values tailored to each demographic–attitudinal segment (gender × age × income × trust × country), as inferred by the LLM. The order of the three decoy-based choice situations for each flight type was randomly determined. Attention check refers to the choice situation in which one option strictly dominates the other two in both price and emission offset. Participants who did not choose this dominant option at least once were excluded from the analysis. The median completion time was approximately 10 minutes. All parameters for decoy options were identified before conducting the experiment and preregistered.

To quantify how LLM-informed decoys and individual characteristics influence participants' offsetting choices, we estimated a binary discrete choice model. The analysis focuses on choices between the target and competitor options, allowing us to isolate preference shifts toward offsetting while excluding situations in which the decoy itself is chosen. Let $y_{in}$ denote the choice outcome of participant $i$ in choice situation $n$ as

$$y_{in} = \begin{cases} 1, \text{if target is chosen;} \\ 0, \text{if competitor is chosen.} \end{cases} \tag{1}$$

All observations in which the decoy option is chosen are excluded from the estimation. The probability of choosing the target option is specified as:

$$P(y_{in} = 1) = \frac{1}{1+\exp(-V_{in})}, \tag{2}$$

where $V_{in}$ is given by:

$$\begin{aligned} V_{in} = \; & b_0 + b_1 * Age_i + \; b_2 * Income_i + b_3 * Gender_i + b_4 * Concern_i + b_5 * Trust_i + \\ & + b_6 * \Delta Price_{in} + b_7 * CountryOptimal_{in} + \; b_8 * CountryNonOptimal_{in} + \\ & + b_9 * IndividualOptimal_{in} \end{aligned} \tag{3}$$

In Equation 3, *Age*, *Income*, *Gender*, *Concern*, and *Trust* are individual-level binary covariates, with reference categories set to "lower than median" for Age and Income, "man" for Gender, and "No" for Concern and Trust. *ΔPrice* is a continuous variable defined as the price of the target option minus the price of the competitor option. The decoy type is represented by three dummy variables indicating whether the participant is exposed to a country-optimal, country-non-optimal, or individual-optimal decoy. The no-decoy condition serves as the reference category. Consequently, coefficients $b_{7-9}$ in the Equation 3 estimate the change in probability of choosing the target option associated with each decoy condition relative to the no-decoy baseline.

We estimated this model using a generalized linear model (GLM) with a binomial family and logit link function using *statsmodels* library in *python* using maximum likelihood. To account for repeated observations per respondent, standard errors were computed using a cluster-robust covariance estimator with clustering at the respondent level.

Based on the estimated model coefficients, we tested two hypotheses regarding the effectiveness of the LLM-informed decoys. First, we tested whether the country-optimal decoy generated a stronger decoy effect than the country non-optimal decoy, i.e., $b_7 > b_8$. Second, we assessed whether personalization improved performance by testing whether the individual-optimal decoy effect exceeded the country-optimal effect, i.e., $b_9 > b_7$. These hypotheses were evaluated using Wald tests, with the significance level set at .05.

### *LLM-Human Alignment*

To investigate the effectiveness of LLM-informed priors in choice models, we employed the model proposed by Rooderkerk et al., (2011). This model is grounded in the random utility maximization framework, and accounts for the decoy effect by adding a contextual term in the standard utility specification. We made two adaptations to this model specification. First, whereas the original model incorporated separate terms to simultaneously capture attraction, similarity, and compromise effects, we only included the attraction term, as we focused exclusively on attraction decoys. Second, to account for demographic heterogeneity, we allowed the attraction coefficient to vary across participants as a function of demographic characteristics and attitudes.

With these modifications, the systematic utility derived by the participant $i$ from choosing alternative $j$ in the choice situation $n$ is defined as

$$V_{inj} = \beta_{price}\text{Price}_{inj} + \beta_{offset}\text{Offset}_{inj} + (\beta_{attraction} + \mathbf{D}_i\boldsymbol{\gamma}_{att})\text{ATT}_{inj}, \quad (4)$$

where the third term accounts for the attraction effect, $\boldsymbol{\gamma}_{att}$ is the vector of estimated coefficients, and $\mathbf{D}_i$ is a vector of demographic variables and attitudes:

$$\mathbf{D}_i = \left(\text{Woman}_i,\ \text{Old}_i,\ \text{HighIncome}_i,\ \text{TrustYes}_i,\ \text{ConcernYes}_i\right). \tag{5}$$

In Equation 4, $\text{ATT}_{inj}$ is calculated based on the distance between alternatives in the attribute space (see Appendix E). A positive $\text{ATT}_{inj}$ means that the alternative gains attraction because it dominates another alternative in the current choice situation, whereas a negative value means that it is dominated by another alternative.

Finally, assuming Gumbel-distributed independent and identically distributed error terms in the indirect utility, the corresponding probability that the participant $i$ chooses alternative $j$ in the choice situation $n$, i.e., $y_{in} = j$ is:

$$P(y_{in} = j \mid \mathbf{X}_{in}, \mathbf{D}_i, \boldsymbol{\beta}_0, \boldsymbol{\gamma}_{att}) = \frac{\exp(V_{inj})}{\sum_{k\in\{C,T,D\}} \exp(V_{ink})}. \tag{6}$$

where $\mathbf{X}_{in}$ includes price, offset, and the attraction term, and $\mathbf{D}_i$ denotes the demographic characteristics of $i$-th traveler, such as gender, age, income, trust, and concern, and

$$\boldsymbol{\beta}_0 = \left(\beta_{offset}, \beta_{price}, \beta_{attraction}\right), \boldsymbol{\gamma}_{att} = (\gamma_{gender}, \gamma_{age}, \gamma_{income}, \gamma_{concern}, \gamma_{trust}) \tag{7}$$

are the estimated coefficients.

We estimated the model using a Bayesian approach under two prior specifications: uninformative priors and LLM-informed priors. In the uninformative-prior condition, all coefficients receive broad Gaussian priors:

$$\beta_m \sim \mathcal{N}(0, 100^2),\ m \in \{offset, price, attraction\}, \tag{8}$$

$$\gamma_l \sim \mathcal{N}(0, 100^2), \qquad l \in \{gender, age, income, concern, trust\}.$$

In the LLM-prior condition, priors are centered on the estimates that we obtain by estimating this model on the LLM-based responses.

$$\beta_m \sim \mathcal{N}(\mu_m^{LLM}, (\xi s_m^{LLM})^2),\ m \in \{offset, price, attraction\}, \tag{9}$$

$$\gamma_l \sim \mathcal{N}(\mu_l^{LLM}, (\xi s_l^{LLM})^2),\ l \in \{gender, age, income, concern, trust\},$$

where $\xi$ is a scaling factor applied to the standard deviations of the LLM-informed priors. We considered $\xi = \{1, 10, 100, 1000\}$.

Estimation was performed in PyMC using the No-U-Turn Sampler (NUTS). Two Markov chains were run in parallel, each with 1,000 tuning iterations followed by 1,000 posterior draws. The target acceptance rate was set to 0.95.

To compare fitting and predictive performance across models, we estimated log-likelihood, Brier score and discrepancy between predicted and observed relative share of choosing target option on the test set.

The log-likelihood is calculated as follows:

$$\text{LL} = \sum_{n=1}^{N} \log P(y_n = j_n^*), \tag{10}$$

where $P(y_n = j_n^*)$ denotes the model predicted probability of the alternative that is chosen in the choice situation $n$ and summarized across all choice situations *N*. The less negative log-likelihood values indicate better model performance because a model with higher predictive accuracy assigns larger probabilities to the observed choices, resulting in logarithmic values closer to zero.

The Brier score is calculated as follows:

$$\text{Brier} = \frac{1}{N} \sum_{n=1}^{N} \sum_{j=1}^{J} \left(P_{nj} - Y_{nj}\right)^2 \tag{11}$$

where $P_{nj}$ denotes the predicted probability of choosing alternative $j$ for choice situation $n$, and $Y_{nj}$ denotes the observed choice indicator, where $Y_{nj} = 1$ if alternative $j$ was chosen and $Y_{nj} = 0$ otherwise. Lower Brier score values indicate better predictive accuracy. The Brier score measures squared discrepancies between predicted choice probabilities and observed choice indicators across all alternatives. Lower values indicate better predictive performance. Like the logarithmic score, it is a strictly proper scoring rule (Gneiting and Raftery, 2007).

The Predicted Relative Share of Target (PRST) measures the model-predicted preference for the target option relative to the competitor option (Berkowitsch et al. 2014). This metric is computed for each participant using all trials they completed,

$$\mathrm{PRST}_i = \frac{\sum_{n=1}^{N_i} P_{n,T}}{\sum_{n=1}^{N_i} P_{n,T} + \sum_{n=1}^{N_i} P_{n,C}} \tag{12}$$

where $i$ indexes participants, $n$ indexes the choice situations, $N_i$ is the number of choice situations completed by the $i$-th participant, $P_{n,T}$ and $P_{n,C}$ denote the model-predicted probabilities of choosing the target and competitor options, respectively, in trial $n$.

Similarly, the Relative Share of Target (RST) that is observed is calculated based on the $i$-th participant's choices as

$$\mathrm{RST}_i = \frac{\sum_{n=1}^{N_i} \mathbb{1}(y_{in}=T)}{\sum_{n=1}^{N_i} \mathbb{1}(y_{in}=T) + \sum_{n=1}^{N_i} \mathbb{1}(y_{in}=C)}, \tag{13}$$

where $i$ indexes participants, $n$ indexes choice situations, $N_i$ is the number of choice situations completed by the $i$-th participant, $y_{in}$ is the observed choice of the $i$-th participant in the $n$-th choice situation, $\mathbb{1}(\cdot)$ is the indicator function.

Finally, the reported discrepancy between the predicted and observed relative share of choosing target is calculated as

$$|\mathrm{PRST} - \mathrm{RST}| = \frac{1}{I} \sum_{i=1}^{I} |\mathrm{PRST}_i - \mathrm{RST}_i| \tag{14}$$

where $I$ is the number of travelers. Lower values of this metric indicate better predictive performance because they reflect smaller discrepancies between predicted and observed target shares. In contrast to log-likelihood or the Brier score, this measure directly evaluates the model's ability to capture variations in relative preference induced by different decoy configurations.

# Results

## *Results of LLM-based simulations*

To test whether the LLM was nudged by a decoy to choose the target, we considered the change in the offsetting probability under different decoy parameters (Fig. 3a). Across all five countries, the LLM showed that decoys priced above the carbon-neutral ticket (the target) and offering lower carbon-offset levels yielded the largest increase in offsetting probability.

Figure 3: Change of the LLM-inferred Offsetting Probability under Country-specific Decoys (a) and overlap between Country-specific and Individual-specific decoys (b)

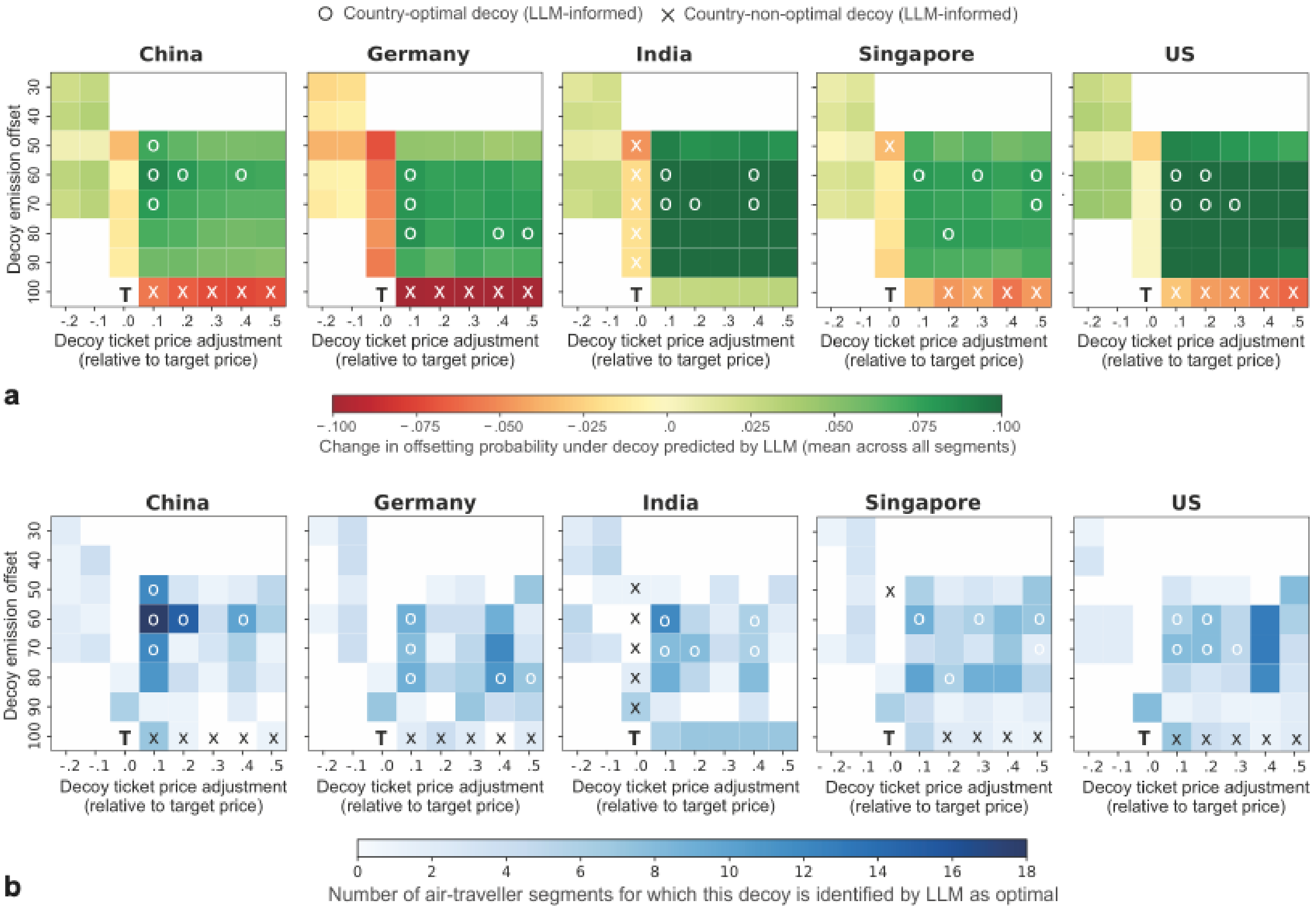

*Notes*: Heatmaps reflect LLM-generated results and display the predicted changes in offsetting probability (a) and the number of air-traveler segments showing maximal increase in offsetting probability (b) as a function of decoy parameters (decoy ticket price adjustment (μ) and decoy emission offset). The ticket prices for the decoy were set as the "carbon-neutral price plus μ*(carbon-neutral price – standard price)". The carbon-neutral ticket (the target) is shown as "T" on the x-axis. Cell color indicates the mean change in offsetting probability (averaged across all air-traveler segments in each country) when the decoy is introduced (red - decrease; green - increase). The cell is overlaid with "○" for "optimal" decoys (top 5 of change in magnitude) or "✕" for "non-optimal" decoys (bottom 5 of change in magnitude).

These country-optimal decoys (marked with “○”) resulted in an average increase of approximately 5-10 percentage points in the probability of full offsetting compared to the no-decoy baseline. Conversely, decoys that were inferior along only one dimension (e.g., same price but lower offset, or same offset but higher price than carbon-neutral ticket) reduced offsetting probability across all countries and were considered as country-non-optimal decoys (marked with “╳”). Finally, decoys priced slightly lower than the target but offering substantially lower emission offsets resulted in a moderate (~2%) increase in offsetting probability in China, India, Singapore, and the United States, while producing no change in offsetting probability in Germany.

To examine whether country-optimal and country non-optimal decoys overlap with individual-optimal decoys, we consider the number of traveler segments for which each decoy is identified by the LLM as optimal (Fig. 3b). The results show that country-optimal decoys frequently correspond to decoy parameter combinations that are optimal for many segments, as indicated by the dense coloring of cells marked with “○”. At the same time, some decoys that are optimal for many segments do not align with country-optimal decoys. For example, in the United States, the LLM identifies decoys as optimal for most segments when their price is set to the carbon-neutral price plus 40% of the difference between the carbon-neutral and standard prices, as reflected by the dense coloring of these cells. Notably, these decoys are not classified as country-optimal (i.e., not marked with “○”), and the prices of country-optimal decoys lie below this level. Finally, partial overlap is also observed between LLM-informed segment-specific decoys and country non-optimal decoys, as indicated by several non-zero colored cells marked with “╳” across all countries.

Taken together, the results of the LLM-based simulations show that LLMs, similar to humans, may change their preferences in the presence of a decoy. We observe heterogeneity in the decoy effect strength and direction across decoy parameters and identify decoys that maximize preference shifts toward offsetting at the country level (country-optimal), as well as decoys that produce a smaller effect (country non-optimal). We further observe heterogeneity in optimal decoy parameters across demographic and attitudinal segments, offering the possibility of increasing the overall effectiveness of decoy nudges through targeted segment-specific parameter tailoring.

### *Results of Field Experiment with Air Travelers*

To assess the effectiveness of the LLM-informed decoys, we analyzed data from 3,495 respondents who completed the survey and passed all attention and consistency checks (see Appendix F). The results revealed that country-optimal decoys increased the probability of choosing carbon offsetting relative to the no-decoy baseline in Singapore, Germany, the United States, and India, but not in China. Individual-optimal decoys, in turn, increased the offsetting probability relative to the no-decoy baseline in Singapore and Germany but did not produce an increase in the other countries (see Appendix G). Our results further showed that country-optimal decoys produced a significantly greater increase in offsetting probability than country-non-optimal decoys only in the United States. In Germany, a similar tendency emerged, although the difference did not reach statistical significance (see Table 1). In all other countries, we observed no significant difference between country-optimal and country-non-optimal decoys. When comparing individual-optimal and country-optimal decoys, we found no significant difference in Singapore or Germany. In the other countries, however, individual-optimal decoys reduced the offsetting probability relative to country-optimal decoys. Thus, our results support the hypothesis that the LLM-informed optimal decoy outperforms the non-optimal decoy only in the United States, but not in the other countries. Moreover, the hypothesis that LLM-informed individually optimal decoys outperform country-level optimal decoys is not supported in any country.

Table 1. Results of the comparison of the effects of LLM-informed optimal and non-optimal decoys on offsetting probability in the human sample. Statistical significance is assessed using Wald tests of differences between the estimated regression coefficients.

| | Singapore | Germany | US | India | China |
|---|---|---|---|---|---|
| Country-optimal ($\boldsymbol{b_7}$) vs.<br>Country non-optimal ($\boldsymbol{b_8}$) | z =.633<br>p = .527 | z = 1.913<br>p = .056 | z = 2.509<br>p = .012* | z = -1.092<br>p = .275 | z = .898<br>p = .369 |
| Individual-optimal ($\boldsymbol{b_9}$) vs<br>Country-optimal ($\boldsymbol{b_7}$) | z = 1.274<br>p = .203 | z = .322<br>p = .748 | z = -2.581<br>p = .010* | z = -4.031<br>p <.001*** | z = -2.129<br>p = .033* |
| Number of respondents | 694 | 638 | 736 | 714 | 713 |
| Number of choice situations | 2776 | 2552 | 2944 | 2856 | 2852 |

To further investigate the limited effectiveness of LLM-informed individual-level decoys, we analyzed the distribution of differences in offsetting probability between individual-optimal and country-optimal decoys across traveler segments. Table 2 presents the number of segments exhibiting positive, negative, or zero changes in offsetting probability, together with the ranges of the observed positive and negative effects.

Across all countries, individual-optimal decoys increased offsetting probability for some segments, whereas other segments showed either no change or even a reduction in offsetting probability. Moreover, among the segments showing positive effects, the effect sizes varied substantially.

Table 2. Number of segments exhibiting positive, negative, or zero changes in offsetting probability between individual-optimal and country-optimal LLM-informed decoys. The ranges of the observed changes in offsetting probability across segments are also shown.

| | Number of segments | | | Change in probability of choosing target | |
|---|---|---|---|---|---|
| | Positive effect | Zero effect | Negative effect | Positive effect | Negative effect |
| Singapore | 13 | 15 | 8 | .001 to .222 | -.074 to -.033 |
| Germany | 14 | 12 | 7 | .002 to .167 | -.033 to -.004 |
| India | 8 | 12 | 13 | .003 to .333 | -.900 to -.009 |
| US | 10 | 11 | 10 | .008 to .200 | -.357 to -.001 |
| China | 2 | 16 | 8 | .004 to .004 | -.043 to -.007 |

Taken together, the results of the human experiment indicate that LLMs can capture preference heterogeneity in some traveler segments. However, their predictions remain insufficiently aligned with the actual behavior of other segments, thereby diminishing the overall advantage of LLM-informed individual-optimal decoys at scale.

### *Results of LLM-Human Alignment*

We suggest that choice models estimated using LLM-generated data can provide informative prior distributions for model parameters. However, incorporating human data into the modelling process is necessary to update these priors and improve their alignment with observed behavior.

To this end, we estimated choice models with LLM-informed and uninformative priors (see Methods). We aim to identify the minimum proportion of human data required to achieve the model performance comparable to that of the entire sample. The results are presented in Figure 4 using two predictive performance metrics: the discrepancy between predicted and observed relative share of target (|PRST − RST|) and the Brier score. For both metrics, lower values indicate better predictive performance.

According to Fig. 4a, a substantial difference in |PRST − RST| is observed between the model estimated using the full human dataset (Human) and the model estimated solely from LLM-generated responses (LLM). These differences are more visible in Brier score (Fig. 4b), as it evaluates the entire predicted probability distribution across all available alternatives rather than focusing solely on the resulting target share (see Methods).

The results also show that, when only a small fraction of human data is available (2% and 5% of the full sample), models estimated with LLM-informed priors achieve a lower |PRST − RST|. This finding indicates an improved ability to predict preference shifts induced by the decoy effect and suggests that LLM-generated priors can partially compensate for limited human data. The same conclusion is derived based on the log-likelihood metric (see Appendix H).

Figure 4: Discrepancy between PRST and RST (a) and Brier scores (b) for the choice model estimated using uninformative and LLM-informed priors

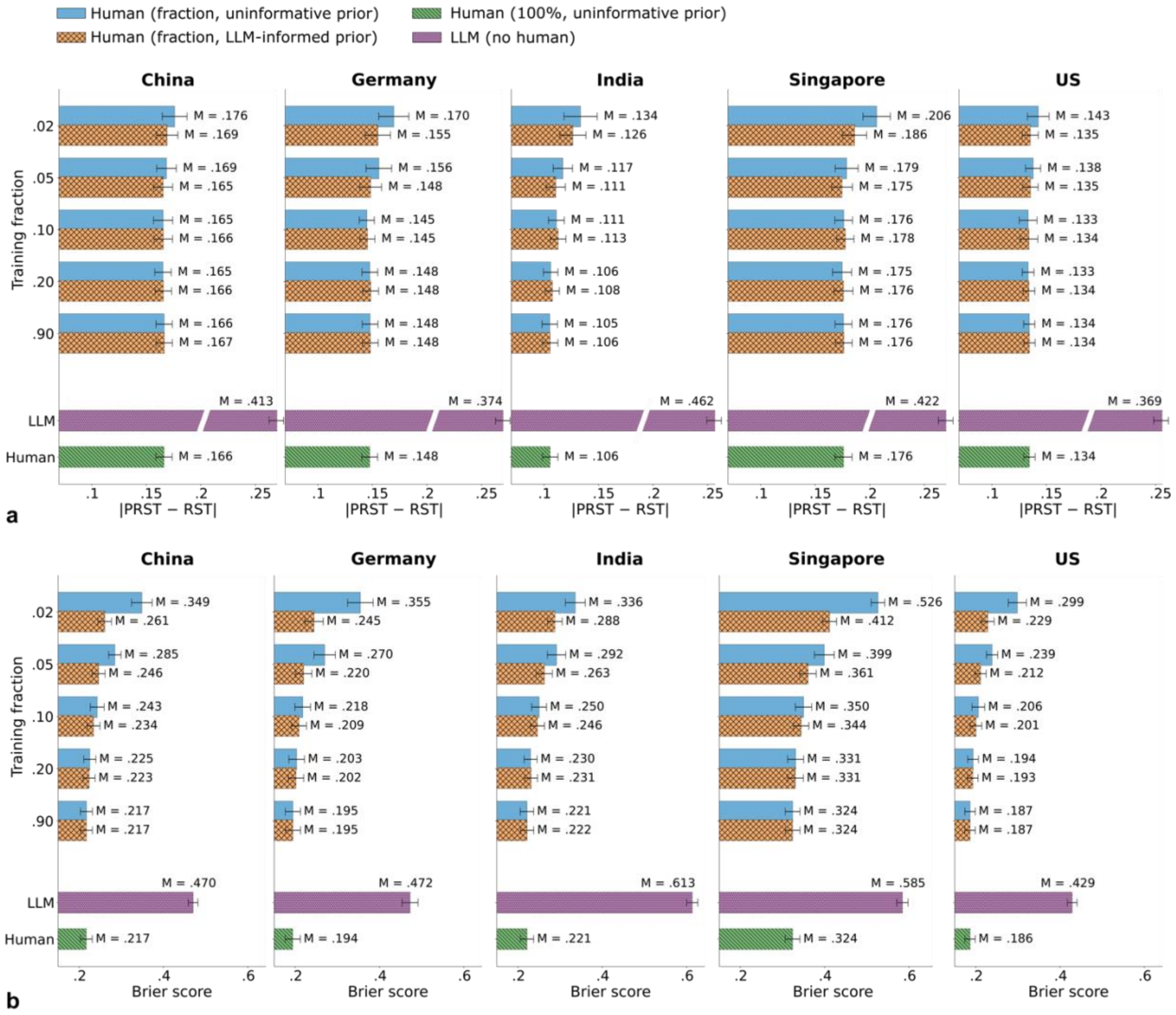

*Notes:* Discrepancy between PRST and RST (a) and Brier scores (b) are shown as mean across train-test splits and 95% confidence intervals. Results include uninformative-prior and LLM-prior models as well as benchmarks (Human and LLM). uninformative-prior and LLM-prior models are estimated using varying fractions of human training data (2%, 5%, 10%, 20%, and 90%), whereas the Human benchmark corresponds to a model estimated using the full human dataset. The LLM condition represents predictions generated directly from LLM-derived parameters without fitting to human data. Results for a particular width of the LLM-prior ($\xi = 100$) are shown. Effect of the width of the LLM-prior is considered in Appendix I.

Analysis of sample size savings reveals that, when using 2% of the human data, the choice model with LLM-informed priors achieves the same Brier score as the model with uninformative priors trained on 4.7–7.9% of the data, depending on the country. This corresponds to savings of 57–74% in human sample size. With 5% of the human data, the savings range from 40% to 49%. A similar pattern is observed for |PRST − RST| (see Appendix K for details).

Finally, we examine how the amount of human training data affects the model parameter estimates. For each training fraction, we calculate the posterior mean for every data split, average these posterior means across splits, and report the corresponding 95% confidence interval (see Fig. 5). Two main insights emerge from Fig. 5. First, substantial differences are observed between the coefficients estimated using solely LLM data and the reference coefficients obtained from the full human dataset. Using LLM estimates as informative priors with a small sample of human data helps stabilize coefficient estimation.

Even with 10% of the human data, substantial differences remain in the coefficient estimates between models with uninformative and LLM-informed priors. By contrast, response-level metrics show much weaker differences: the PRST–RST discrepancy has largely disappeared, and the Brier-score improvement is only marginal. This finding suggests that accurate prediction of aggregate outcomes does not necessarily imply accurate recovery of the underlying model parameters. Second, coefficient estimates obtained with uninformative priors exhibit considerable variability across train–test splits as reflected in large confidence intervals. This variability arises because, when only a limited amount of human data is available, different training samples may contain substantially different proportions of traveler segments. Furthermore, some segments may be underrepresented or entirely absent in a training split. As a result, the estimated segment-specific coefficients become unstable and highly sensitive to the composition of the training data.

Figure 5: Model coefficients estimated using uninformative and LLM-informed priors

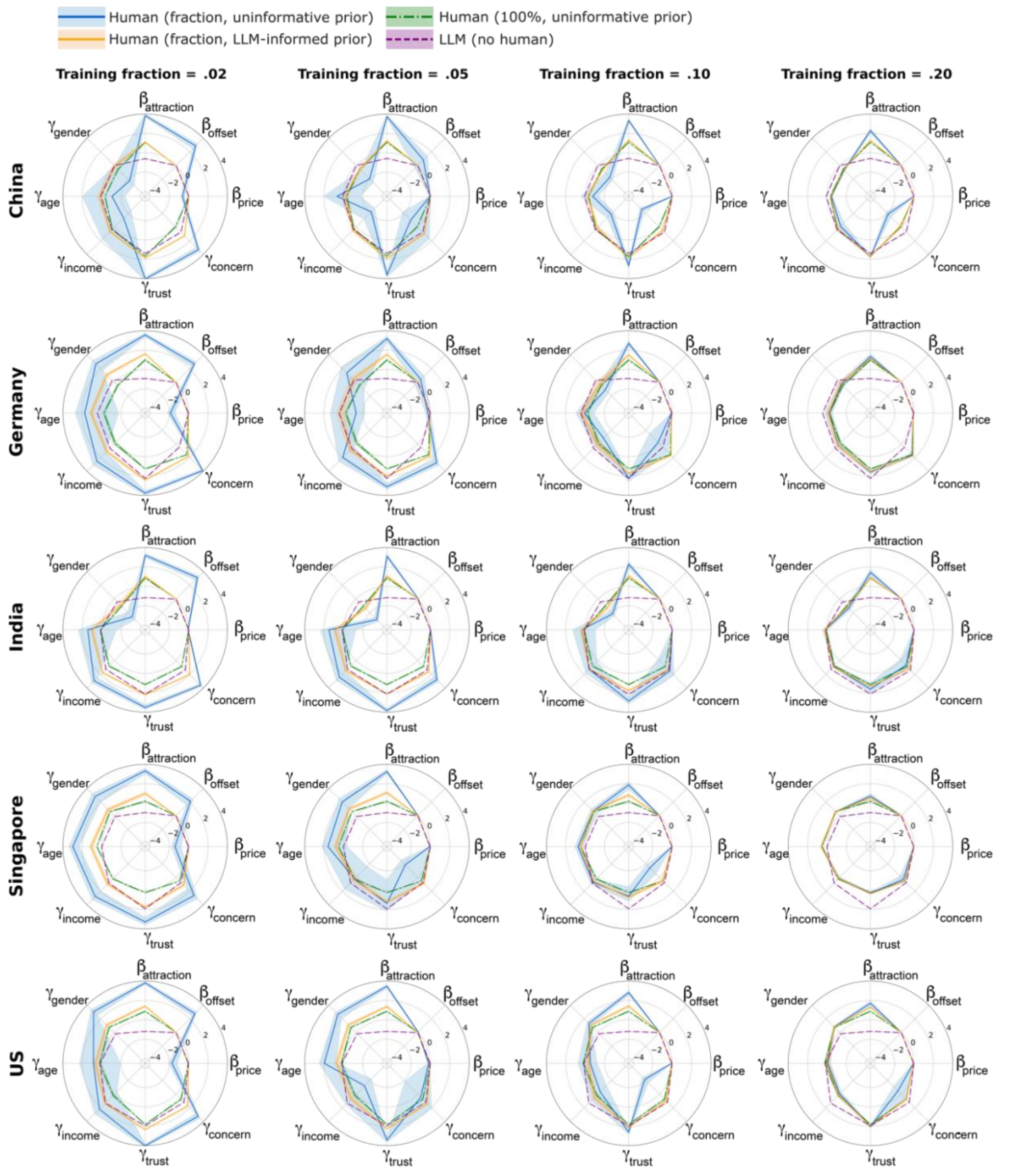

*Note:* Each panel presents the estimated coefficients in log-scale describing the effects of demographic characteristics and attitudinal variables on the context-dependent utility component that captures the decoy effect (see Methods). Coefficients (means of posterior means across splits and 95% credible intervals) are shown for models estimated using uninformative prior, LLM-informed priors, as well as two benchmark models estimated using the full human dataset (Human) and solely LLM-generated responses (LLM). Results for a particular width of the LLM-prior ($\xi = 100$) are shown. Effect of the width of the LLM-prior is considered in Appendix J.

## Discussion and Conclusions

Nudge-based interventions offer a relatively low-cost and non-coercive means of influencing behavior. However, their effectiveness varies across individuals. Designing personalized nudges therefore requires extensive experimentation involving respondents from different population segments. Using a decoy-based nudging strategy for aviation carbon offsetting as a case study, we examine whether an LLM can reduce this experimental burden by identifying promising segment-specific decoy configurations and informing behavioral models when human data are scarce.

Based on simulations with OpenAI's GPT-4o-mini, we show LLMs exhibit decoy-induced preference shifts, and the magnitude of these shifts varies across decoy configurations. To the best of our knowledge, this is the first study to demonstrate that LLMs are susceptible to the decoy effect, a well-established context effect that violates the regularity principle of rational choice. A recent study by Peng et al. (2026) failed to reproduce the decoy effect using LLM-based digital twins. However, that study examined only a limited set of decoy configurations. Given the substantial heterogeneity in effect magnitude across decoy parameters observed in our study, the configurations examined by Peng et al. (2026) may not have been well suited to eliciting the effect in LLMs. Re-evaluating these digital twins across a broader range of decoy parameters would help determine whether nudging effects emerge for other configurations.

The sensitivity of LLMs to the decoy effect has important implications for both policymakers and marketers. LLMs that reproduce context-induced shifts in human preferences may serve as low-cost testbeds for evaluating behavioral interventions before large-scale human experimentation. Moreover, the possibility of influencing an LLM through the composition of a choice set may create new strategic opportunities in future e-commerce environments in which LLM agents act as personal shoppers. As these agents gain greater autonomy in purchasing decisions, marketing strategies may increasingly target the decision processes of the agents themselves rather than those of human consumers. At the same time, this possibility raises important ethical and consumer-protection concerns about the manipulation of AI agents acting on users' behalf.

The human experiment involving respondents across five countries provides only partial validation of the LLM-informed decoy recommendations. Country-optimal decoys informed by LLM produced significantly stronger nudging effects than country-non-optimal decoys in the United States, while Germany showed a similar tendency that did not reach conventional statistical significance. No corresponding advantage was observed in China, India, or Singapore. This pattern indicates that LLMs can identify relatively effective decoy configurations in some national contexts, but the advantage does not generalize consistently across countries. One possible explanation is a Western-centric bias in GPT: the model may be better calibrated to behavioral patterns, consumer contexts, and policy narratives that are more strongly represented in Western data (Atari et al. 2023; Johnson et al. 2022). However, the present study does not establish this mechanism, and the cross-country differences may also reflect variation in traveler preferences, sample composition, or other unobserved cultural and contextual factors. Future research should assess this explanation using multiple LLMs, languages, culturally adapted prompts, and country-specific models.

Individual-optimal decoys did not significantly outperform country-optimal decoys in any country. Although segment-specific decoys increased the probability of offsetting for some traveler segments, they had no effect or even reduce offsetting for others, thereby eliminating any aggregate advantage. These findings indicate that using LLMs as direct substitutes for human respondents to identify personalized decoy parameters is not currently feasible. Instead, we propose a Bayesian approach to reduce the misalignment between LLM-generated and human responses. Within this framework, LLM responses provide informative priors for estimating a choice model with only a small fraction of human data, including settings in which some traveler segments are absent from the human subsample. This approach has the potential to reduce the costs of human experimentation by lowering sample-size requirements and the need to obtain representative coverage of every consumer segment.

Finally, this approach offers a more theory-grounded route to alignment because it estimates interpretable choice-model coefficients rather than aligning observed responses alone. Brand et al. (2023) also fit choice model using LLM's responses and compare willingness-to-pay estimates derived from LLM and human samples. They report substantial discrepancies between the two sets of estimates. When LLMs are fine-tuned using human responses, the resulting willingness-to-pay estimates move closer to the human estimates but remain imperfectly aligned. Our results suggest that, rather than fine-tuning an LLM to mimic human choices directly, a promising strategy may be to use a small amount of human data to update LLM-informed priors through Bayesian estimation, thereby aligning behavior at the level of model parameters rather than observed responses alone.

In their recent work, Wang et al. (2026) also seek to improve estimates of choice-model coefficients by combining human and LLM-generated data. Although their method aligns the behavior of LLMs and humans at the response level, its empirical evaluation focuses on estimating model coefficients. Consistent with our findings, they show that combining limited human data with LLM-generated responses can improve coefficient estimates. At the same time, their model does not capture preference heterogeneity across consumer profiles, and they identify incorporating personas to represent specific populations as a direction for future research. Thus, our study extends this line of research by modeling variation in decoy susceptibility across demographic and attitudinal profiles.

Our study has several limitations. First, we relied solely on GPT-4o-mini for decoy design and did not test whether other models, such as Claude, Gemini, or Mistral, produce similar results. These models differ in architecture, training procedures, and underlying training data. It therefore remains unclear whether our findings are model-specific or more broadly generalizable. Second, the human experiment was conducted in a stated-preference setting, and it remains unclear whether the behavioral effects of the decoys would persist over time or translate into real-world decision-making. Future studies should validate LLM-informed interventions using revealed-preference data. Finally, the psychological mechanisms underlying LLM susceptibility to the decoy effect remain poorly understood. Future research could examine these mechanisms using process-level analyses and controlled manipulations of prompts and model settings.

## *Appendix A: Traveler Segments*

Table A.1: Segments of air travelers used for simulations with LLMs

| **Factor** | **Levels** |
|---|---|
| Country of residence | India, China, Germany, the US, Singapore |
| Gender | man, woman |
| Age | below median age, above median age |
| Income | above median income, below median income |
| Environment protection concern | cannot say that you are concerned, you are concerned |
| Trust to offset programs | cannot say that you believe, believe |

## *Appendix B: Decoy Parameters*

Here we focus on the attraction decoy and define two areas of decoy parameters (see Fig. B.1).

Figure B.1: Illustration of the decoy parameter regions (I and II) in the two-dimensional attribute space defined by ticket price and emission offset.

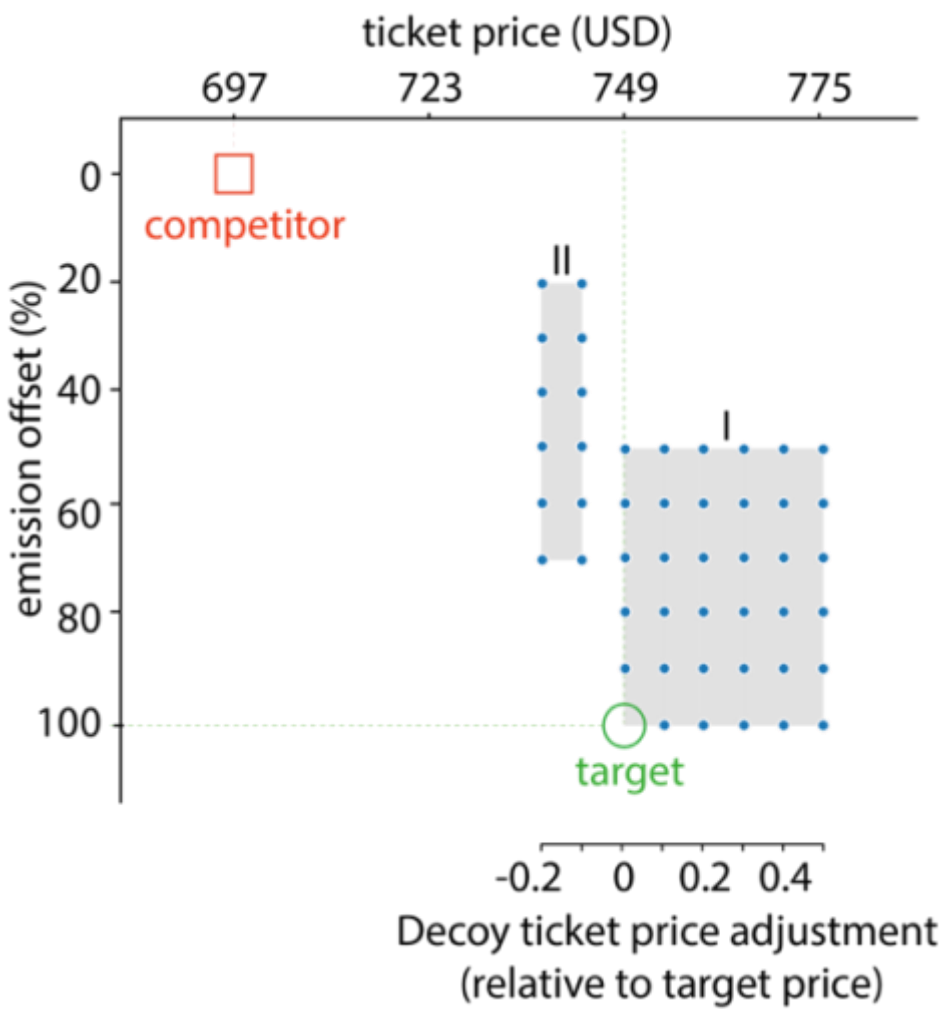

*Notes:* Areas of decoy parameters are shown in gray. The actual 45 decoys that have been tested are indicated by dots.

The area I includes asymmetrically dominated attraction decoys such that the decoy alternatives are fully dominated by the target (worse than target in at least one parameter), similar to Noguchi and Stewart (2018). In this area, the decoy takes 35 discrete values. The ticket prices for the decoy are set as the target price plus μ*(target price – competitor price) where μ is a price adjustment factor taking values 0, .1, .2, .3, .4, .5, and competitor refers to the standard ticket without offset. The area II includes non-dominated decoys which are slightly cheaper than the target but significantly worse in emission offsets, similar to Lichters et al (2015). In this area, the decoy takes 10 discrete values. The price adjustment, μ, takes values -.1 or -.2 meaning that the decoy ticket prices are slightly lower than target price. The offset for the decoy is set as the target offset minus 30%, 40%, 50%, 60%, or 70%. These two areas result in 45 combinations of decoy parameters.

***Appendix C: LLM Prompts***

Table C.1: Structure of the System Prompt for LLM

| |
|---|
| "You are a [man], aged [above median age], permanently residing in [Singapore], and your monthly income is [above median income]. You [concern] environment protection in your daily life and [believe] that the money you pay for carbon offsets are really used to offset emissions." |

Table C.2: Structure of the User Prompt for LLM

| |
|---|
| "You are planning a [flight_length]-hour flight. This flight produces [emissions] kg of $CO_2$ emissions which is equivalent to producing [bottles_number[1]] plastic water bottles. Using all the information below, consider three options: 1. Pay [target_price] [currency] and fully offset emissions; 2. Pay [competitor_price] [currency] and not offset emissions; 3. Pay [decoy_price] [currency] and offset [decoy_offset] emissions. Which option would you choose? Please give your answer only with the option number without any words." |

[1] The number of bottles depends on the amount of $CO_2$ emissions, assuming that producing each bottle generates 97 g of $CO_2$, based on the estimate reported by TAPP Water: https://tappwater.co/blogs/blog/carbon-footprint-bottled-water.

### *Appendix D: Choice Situations*

In each trial, flight duration was randomly drawn from a predefined set corresponding to each category (see Fig. D.1).

Figure D.1: Algorithm for Calculating Competitor, Target and Decoy Parameters.

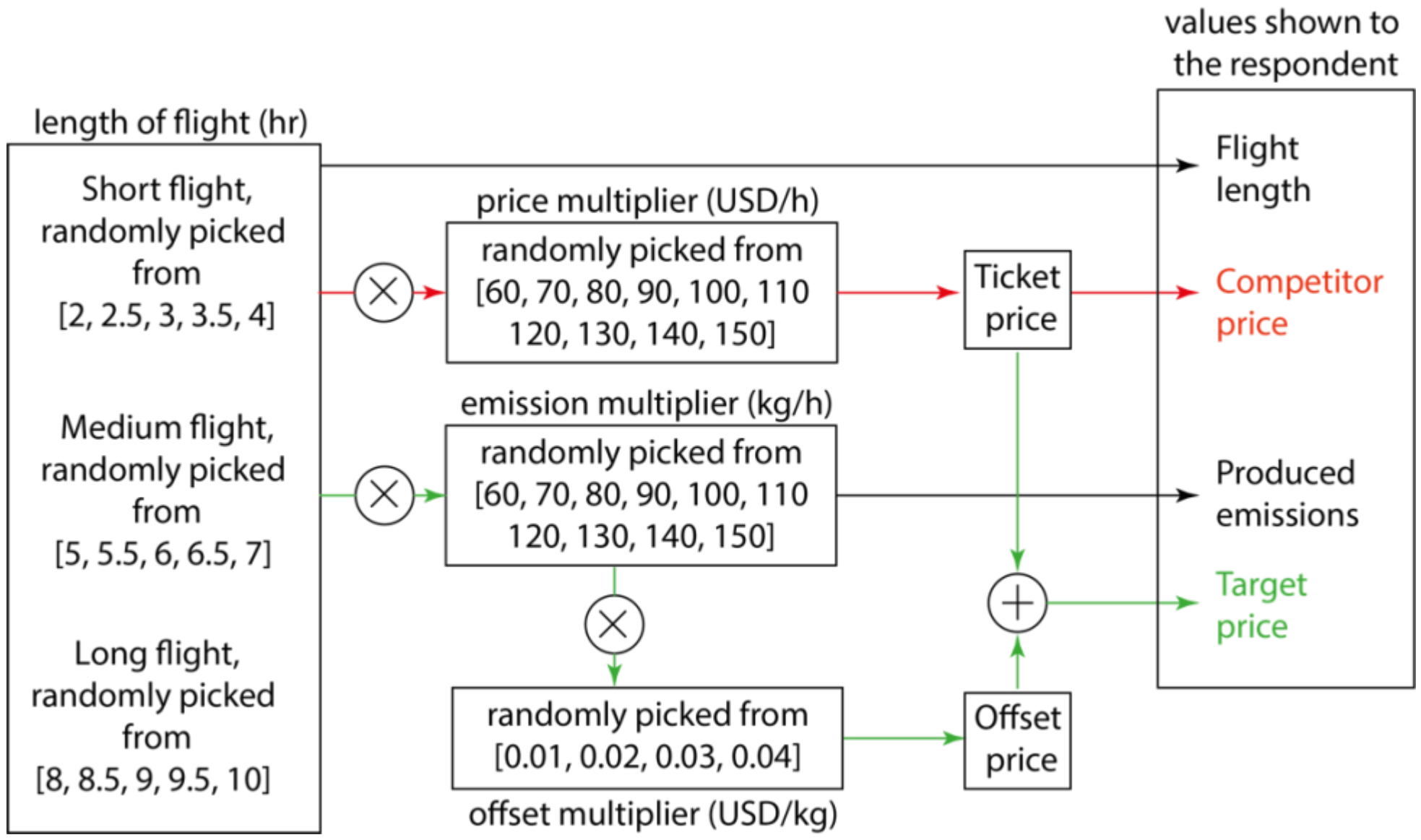

*Notes:* Algorithm for constructing competitor and target ticket prices based on flight length and price, emission, and offset multipliers ('×' denotes multiplication; '+' denotes summation).

The competitor price was computed by multiplying the flight length by a price multiplier, randomly sampled from a discrete range. Importantly, flights of the same duration may result in different ticket prices and carbon emissions due to variations in aircraft type, load factors, routing, and weather conditions. Likewise, identical emission quantities may incur different offset costs depending on the pricing schemes and tariffs applied by offset providers. To capture this variability while maintaining experimental control, both the price multipliers and emission multipliers were randomly sampled from ranges predefined by LLM. Carbon emissions were calculated by multiplying flight length by the emission multiplier, and the offset price was obtained by multiplying total emissions by an offset multiplier, also randomly sampled from an LLM-informed range. The target price was then defined as the sum of the competitor price and the offset price.

### *Appendix E: Calculation of Attraction Term in the Choice Model*

For the participant $i$, alternative $j$ in the choice situation $n$, $\text{ATT}_{inj}$ is calculated based on the distance between alternatives in the attribute space using the following steps. First, we estimate the preference direction in the attribute ranges. Let $a_{1,inj}$ denote the negative value of the price attribute, that is, $-\text{Price}_{inj}$, and $a_{2,inj}$ denote the value of the offset attribute, that is, $\text{Offset}_{inj}$, The preference-direction vector is then defined as:

$$\mathbf{v}_{in} = \left(\max_j a_{1,inj} - \min_j a_{1,inj},\ \max_j a_{2,inj} - \min_j a_{2,inj}\right), \tag{E1}$$

Second, we set unit preference direction as:

$$\mathbf{u}_{in} = \frac{\mathbf{v}_{in}}{\|\mathbf{v}_{in}\|} \tag{E2}$$

We then project each alternative onto this direction by multiplying its attribute values by the corresponding components of the unit preference-direction vector:

$$q_{inj} = a_{1,inj}u_{in,1} + a_{2,inj}u_{in,2}, \tag{E3}$$

Where the maximum possible projection distance within the choice situation is

$$M_{in} = \max_j q_{inj} - \min_j q_{inj}. \tag{E4}$$

Third, we estimate attraction coefficient based on pairwise comparisons between alternatives. When alternative $j$ dominates alternative $k$, the projected distance contributes positively to the attraction of $j$ and negatively to the attraction of $k$:

$$A_{inj} = \sum_{k \neq j} \mathbb{I}_{j \succ k} \left|q_{inj} - q_{ink}\right| - \sum_{k \neq j} \mathbb{I}_{k \succ j} \left|q_{ink} - q_{inj}\right|. \tag{E5}$$

Alternative $j$ dominates alternative $k$ if it is at least as good on both transformed attributes and strictly better on at least one attribute:

$$\mathbb{I}_{j \succ k} = \mathbb{I}\left(a_{1,inj} \geq a_{1,ink},\ a_{2,inj} \geq a_{2,ink},\ \text{and}\ \left[a_{1,inj} > a_{1,ink}\ \text{or}\ a_{2,inj} > a_{2,ink}\right]\right). \tag{E6}$$

Fourth, we normalize attraction coefficient by the maximum projection distance:

$$\text{ATT}_{inj} = \frac{A_{inj}}{M_{in}}, \tag{E7}$$

These values are calculated for each participant, alternative, and choice situation before model estimation.

## *Appendix F: Demography of Respondents*

Table F.1: Demographic Information of survey participants in different countries.

| | India | China | Germany | Singapore | US |
|---|---|---|---|---|---|
| Passed Initial screening | 5363 | 1591 | 1253 | 1532 | 2245 |
| Concerned about environment protection | 97% | 95% | 92% | 89% | 86% |
| Trust in offset programs | 86% | 76% | 61% | 69% | 65% |
| Passed survey | 714 | 713 | 638 | 694 | 736 |
| Men / Women | 352 / 362 | 371 / 342 | 307 / 331 | 347 / 347 | 349 / 387 |
| Above / below median age | 370 / 344 | 362 / 351 | 316 / 322 | 352 / 342 | 342 / 394 |
| Above / below median income | 398 / 316 | 403 / 310 | 331 / 307 | 350 / 344 | 398 / 338 |
| Did not pay offset | 28% | 25% | 39% | 46% | 58% |

*Notes:* Total number of people passed initial screening is 11984 and the total number of people finished survey is 3495. Note that a participant finishes the survey only if they pass three attention checks. In each of these checks, the participant is presented with a choice situation that includes an alternative which outperforms other options in all attributes. We expect that if the participant makes an informed choice based on the provided information, they should choose this alternative. Otherwise, we screen out the participant. Trust and concern are estimated based on people who passed the initial screening, and percentage of those paying offset – based on people who finished the survey. Target group refers to segment predicted by LLM to increase offsetting probability under decoy.

Table F.2. Marginal distribution of age, gender, and income across survey participants finished survey in different countries.

| | Men (%) | Women (%) | Below median age (%) | Above median age (%) | Below median income (%) | Above median income (%) |
|---|---|---|---|---|---|---|
| India | 49.3 | 50.7 | 48.2 | 51.8 | 44.3 | 55.7 |
| China | 52.1 | 47.9 | 49.2 | 50.8 | 43.5 | 56.5 |
| Germany | 48.1 | 51.9 | 50.5 | 49.5 | 48.1 | 51.9 |
| Singapore | 50.0 | 50.0 | 49.3 | 50.7 | 49.6 | 50.4 |
| US | 47.2 | 52.8 | 53.6 | 46.4 | 45.9 | 54.1 |

## *Appendix G: Results of Binary Model on Human Data*

Table G.1: Estimated coefficients $b_1$-$b_9$ from the binary logit model (see Methods: Field Experiment with Travelers) predicting offsetting probability for different decoy conditions in different countries

| **Predictor** | **Singapore** | **Germany** | **US** | **India** | **China** |
|---|---|---|---|---|---|
| Intercept ($\boldsymbol{b_0}$) | .214<br>(p = .537) | .751<br>(p = .085) | 1.767<br>(p <.001***) | .709<br>(p = .036*) | −.013<br>(p = .980) |
| Age ($\boldsymbol{b_1}$):<br>Higher than median<br>(ref. Lower than median) | .351<br>(p = .157) | −.379<br>(p = .152) | −.588<br>(p = .031*) | −.691<br>(p = .019*) | −.093<br>(p = .697) |
| Income ($\boldsymbol{b_2}$):<br>Higher than median<br>(ref. Lower than median) | .215<br>(p = .386) | .043<br>(p = .860) | −.054<br>(p = .827) | .370<br>(p = .213) | 1.205<br>(p = .001**) |
| Gender ($\boldsymbol{b_3}$):<br>Woman<br>(ref. Man) | .587<br>(p = .016*) | −.134<br>(p = .608) | .133<br>(p = .565) | −.248<br>(p = .321) | −.675<br>(p = .008**) |
| Environment concern<br>($\boldsymbol{b_4}$):<br>Yes<br>(ref. No) | .056<br>(p <.001***) | 1.831<br>(p <.001***) | 1.276<br>(p <.001***) | 1.816<br>(p <.001***) | .529<br>(p = .171) |
| Trust in offsetting<br>programs ($\boldsymbol{b_5}$):<br>Yes<br>(ref. No) | .073<br>(p <.001***) | 1.503<br>(p <.001***) | 1.601<br>(p <.001***) | 1.691<br>(p <.001***) | 3.557<br>(p <.001***) |
| Δ Price ($\boldsymbol{b_6}$) | −.025<br>(p <.001***) | −.046<br>(p <.001***) | −.040<br>(p <.001***) | −.0002<br>(p = .001***) | −.0016<br>(p = .049*) |
| Country-optimal<br>decoy ($\boldsymbol{b_7}$)<br>(ref. No decoy) | .120<br>(p = .021*) | .260<br>(p <.001***) | .247<br>(p <.001***) | .637<br>(p <.001***) | .083<br>(p = .096) |
| Country non-optimal<br>decoy ($\boldsymbol{b_8}$)<br>(ref. No decoy) | .094<br>(p = .096) | .156<br>(p = .009**) | .102<br>(p = .134) | .750<br>(p <.001***) | .047<br>(p = .349) |
| Individual-optimal<br>decoy ( $\boldsymbol{b_9}$)<br>(ref. No decoy) | .168<br>(p = .001**) | .277<br>(p <.001***) | .046<br>(p = .620) | .161<br>(p = .192) | −.005<br>(p = .923) |
| Number of respondents | 694 | 638 | 736 | 714 | 713 |
| Number of choice situations | 2776 | 2552 | 2944 | 2856 | 2852 |

*Notes: Age*, *Income*, *Gender*, *Concern*, and *Trust* are individual-level binary covariates, with reference categories set to "lower than median" for Age and Income, "man" for Gender, and "No" for Concern and Trust. *Δ Price* is a continuous variable defined as the price of the target option minus the price of the competitor option. The decoy type is represented by three dummy variables indicating whether the participant is exposed to a *Country-optimal*, *Country-non-optimal*, or *Individual-optimal* decoy. The *No-decoy* condition serves as the reference category.

## *Appendix H: Effect of LLM-informed Priors on Log-likelihood Estimation*

Figure H.1: Log-likelihood for the choice model estimated using uninformative and LLM-informed priors

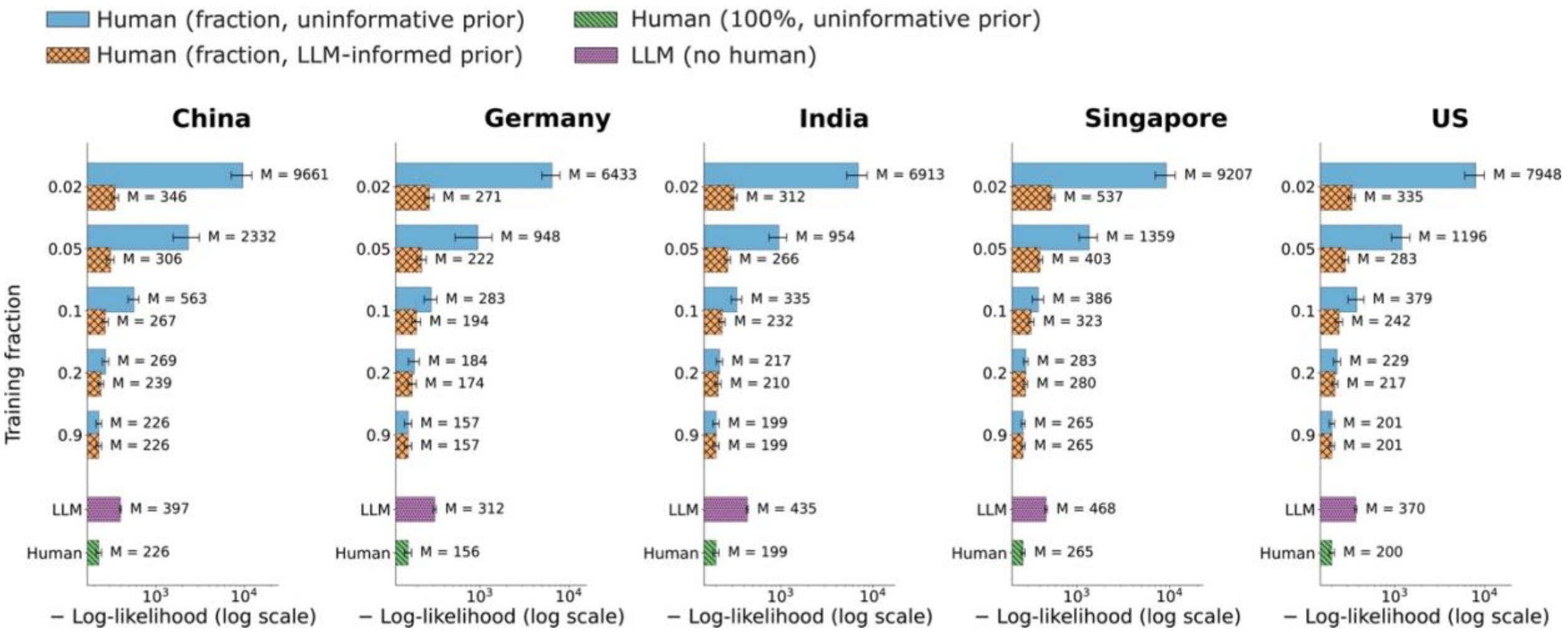

*Notes:* values are shown as mean across train-test splits and 95% confidence intervals. Results include uninformative-prior and LLM-prior models as well as benchmarks (Human and LLM). uninformative-prior and LLM-prior models are estimated using varying fractions of human training data (2%, 5%, 10%, 20%, and 90%), whereas the Human benchmark corresponds to a model estimated using the full human dataset. The LLM condition represents predictions generated directly from LLM-derived parameters without fitting to human data. Results for a particular width of the LLM-prior ($\xi = 100$) are shown. Effect of the width of the LLM-prior is considered in Appendix J.

## *Appendix I: Results stability against width of the LLM-informed prior*

Figure J=I.1: Effect of the width of the LLM-informed prior on PRST-RST (a), Brier-score (b), and Log-likelihood (c) estimated using 2% fraction of human data

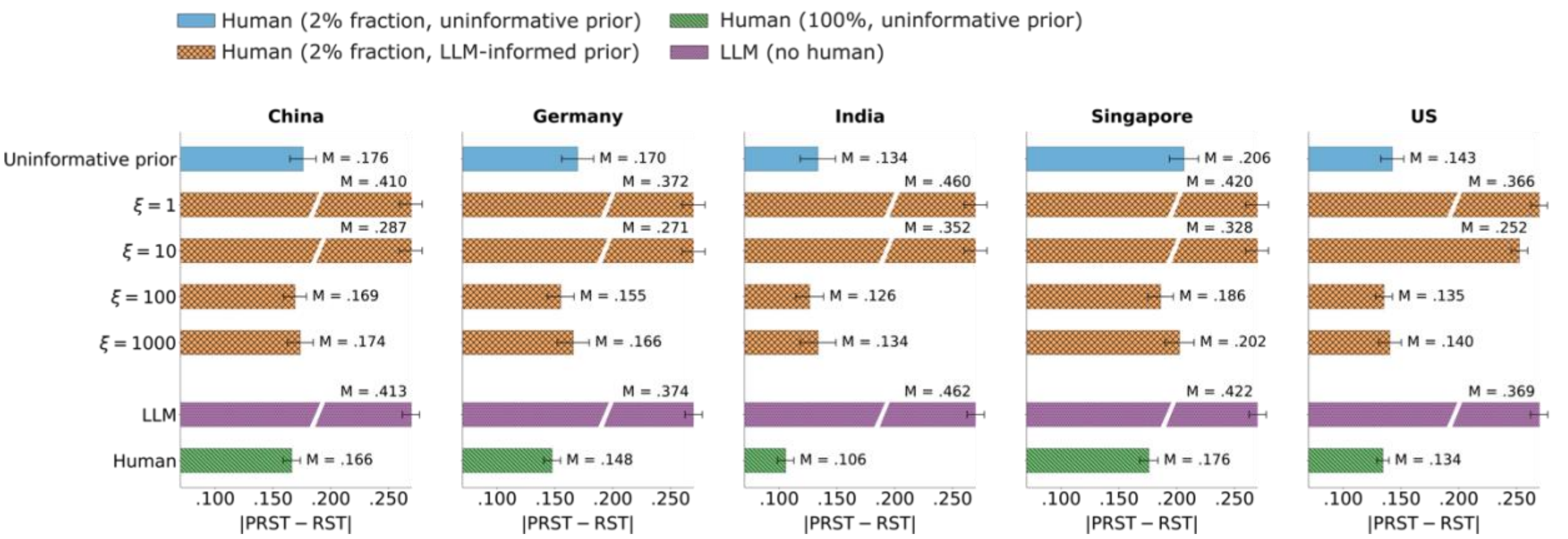

**a**

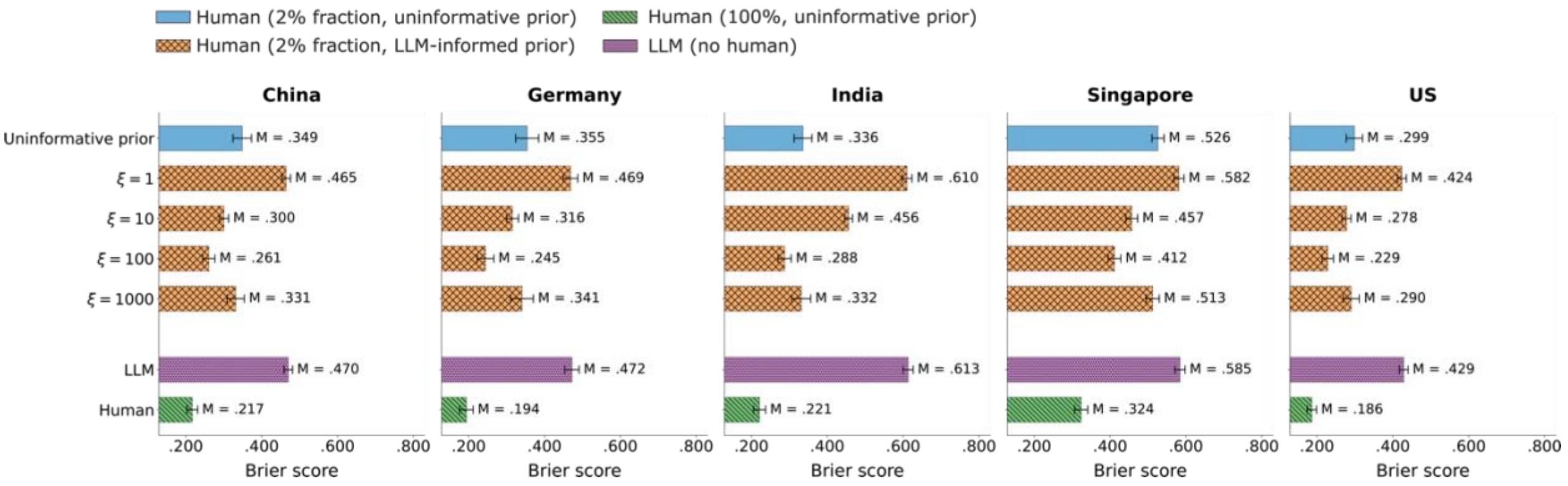

**b**

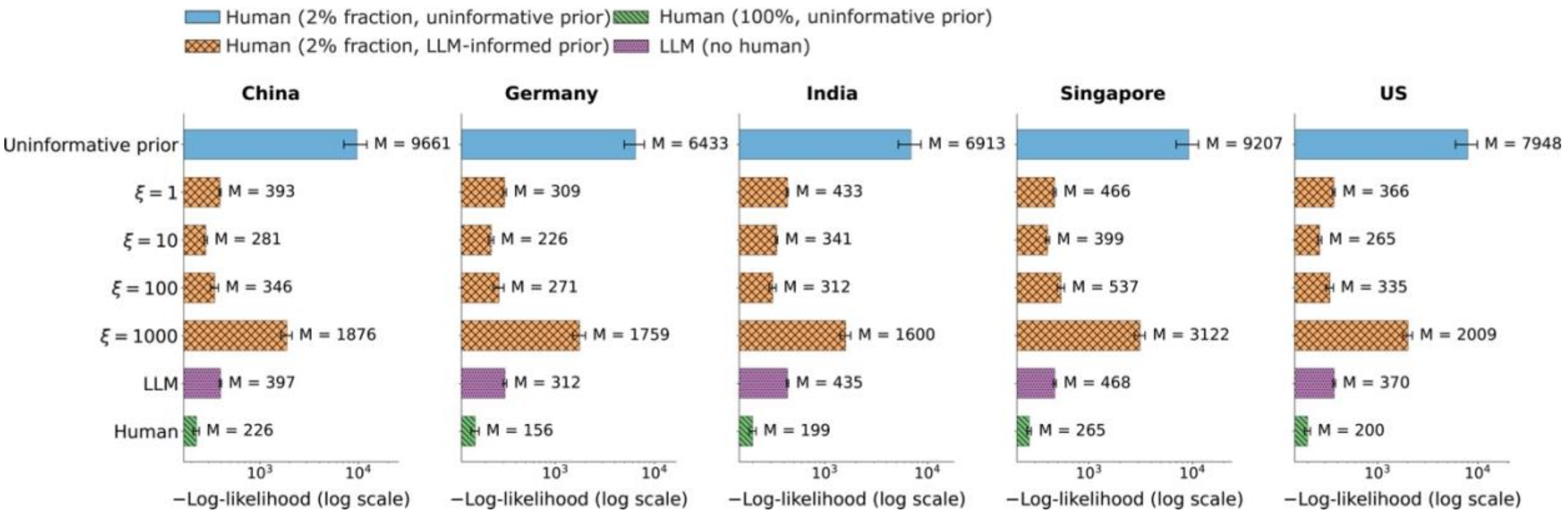

**c**

*Notes:* values are shown as mean across train-test splits and 95% confidence intervals. Results include uninformative-prior and LLM-prior models as well as benchmarks (Human and LLM). uninformative-prior and LLM-prior models are estimated using 2% of human training data, whereas the Human benchmark corresponds to a model estimated using the full human dataset. The LLM condition represents predictions generated directly from LLM-derived parameters without fitting to human data. $\xi$ defines the width of the LLM-informed prior (see Methods)

## *Appendix J: Stability of Coefficient Estimates against Width of the LLM-informed Prior*

Figure J.1: Effect of the width of the LLM-informed prior on model coefficients estimated using 2% fraction of human data

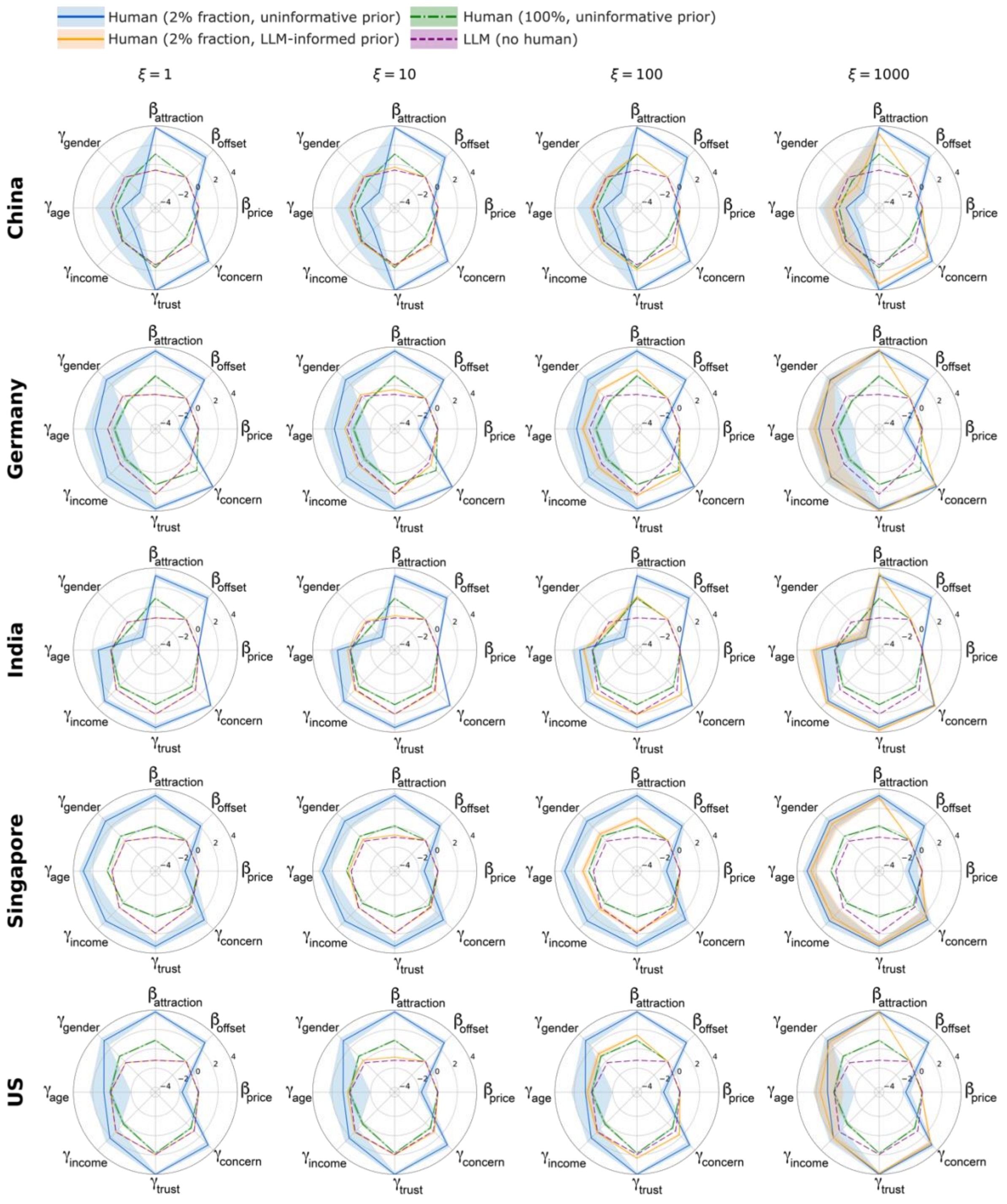

*Note:* Coefficients (means of posterior means across splits and 95% credible intervals) are shown for models estimated using uninformative prior, LLM-informed priors for 2% human data fraction, as well as two benchmark models estimated using the full human dataset (Human) and solely LLM-generated responses (LLM).

## *Appendix K: Estimating human data savings with LLM-informed priors*

We compare how much human training data models with uninformative priors and LLM-informed priors need to achieve the same Brier score and |PRST-RST| on the test set.

***Comparison procedure***

First, we select the mean performance scores (Brier score or |PRST-RST|) for the model with LLM-informed priors at a chosen human training fraction, such as 2%.

Second, we find two adjacent training fractions where the human model with uninformative priors shows performance scores above and below the selected score of the model with LLM-informed priors.

Third, we assume the score for the model with uninformative priors changes linearly between those two adjacent training fractions. Therefore, we obtain the fraction of human data for which the model with uninformative priors achieves the score that matches the selected LLM-prior measure at the 2% training fraction.

$$f_H = f_a + \frac{e_a - e_L}{e_a - e_b} \times (f_b - f_a) \quad \text{(K1)}$$

Here $f_a$ and $f_b$ are the two surrounding fractions of the model with uninformative priors, with scores $e_a$ and $e_b$. The selected score of the model with LLM-informed priors is $e_L$, and $f_H$ is the matching fraction of the model with uninformative priors.

Fourth, we compare the obtained fraction for the model with uninformative priors with the fraction of the model with LLM-informed priors.

$$Savings = (1 - \frac{f_L}{f_H}) \times 100\% \quad \text{(K2)}$$

The fraction used with LLM priors is $f_L = 2\%$. The denominator is the human data fraction needed for the model with uninformative prior to achieve matching score.

Figure K.1. Illustrative example of estimating human sample size savings using Brier scores in China.

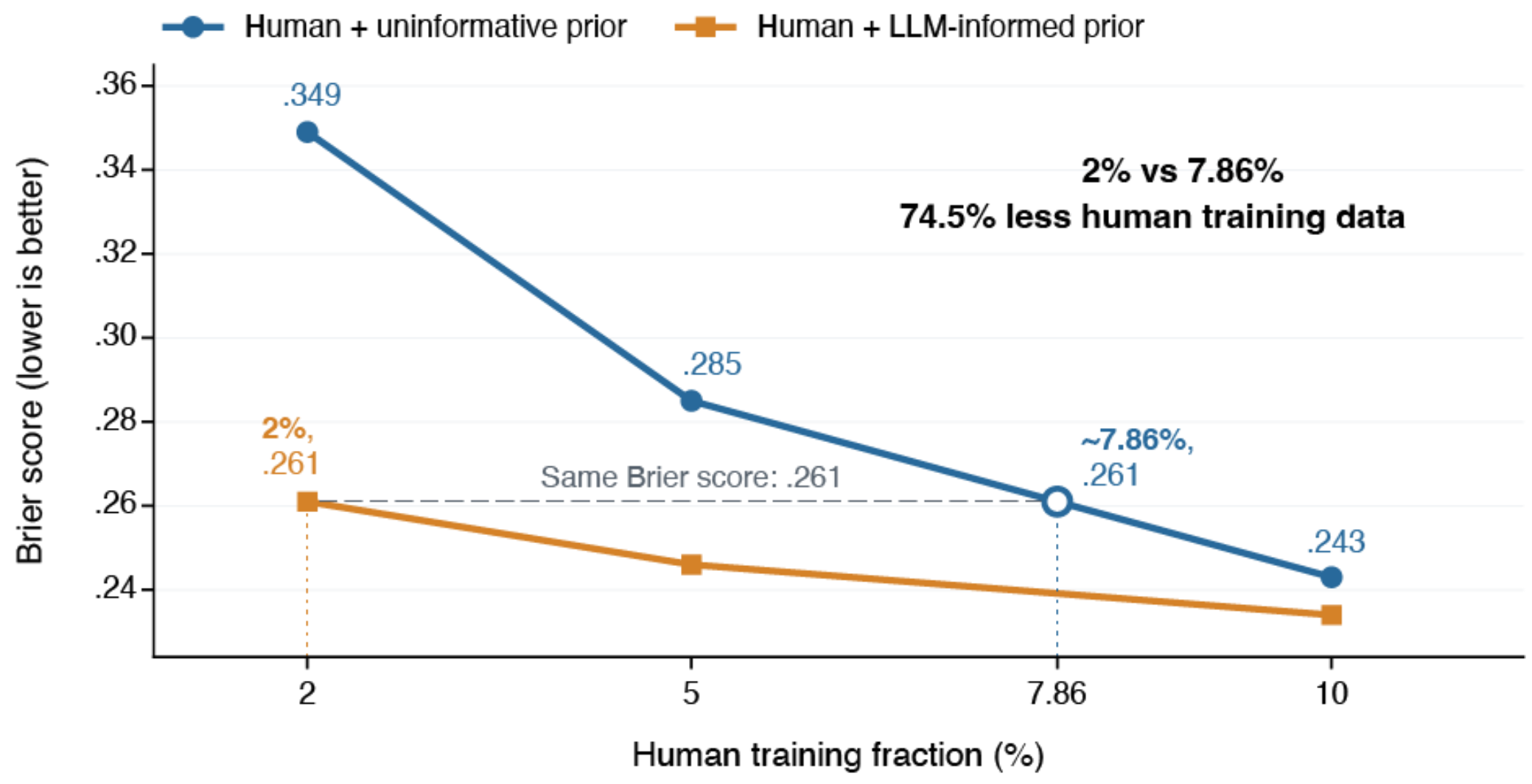

*Notes:* Matching the Brier score horizontally identifies the corresponding required human-only fraction. The point at 7.86% is interpolated.

### *Worked example for China*

With 2% of the human training data, the model with LLM-informed priors has a mean Brier score of .261 (see Fig. 4B). The Brier scores obtained with the model with uninformative priors around this value are .285 at 5% and .243 at 10%. We interpolate between these two human-only points.

$$f_H = 5\% + \frac{.285 - .261}{.285 - .243} \times (10\% - 5\%) \approx 7.857\%$$

$$Savings = (1 - \frac{2\%}{7.857\%}) \times 100\% \approx 74.5\% \qquad \text{(K3)}$$

### *Estimated savings across countries*

For each country, we repeat the same procedure at the 2% and 5% human training levels. The tables below show the matched fractions required for the model with uninformative priors and the resulting relative savings. Percentages refer to the training fractions used in Figure 4.

*Table K.1. Estimated human data savings from LLM-informed priors based on Brier scores*

| | **Model with LLM-informed priors uses 2% of the human sample** | | **Model with LLM-informed priors uses 5% of the human sample** | |
|---|---|---|---|---|
| | Fraction of the human sample required for the model with uninformative priors | Sample size savings | Fraction of the human sample required for the model with uninformative priors | Sample size savings |
| China | 7.86% | 74.5% | 9.64% | 48.1% |
| Germany | 7.40% | 73.0% | 9.81% | 49.0% |
| India | 5.48% | 63.5% | 8.45% | 40.8% |
| Singapore | 4.69% | 57.4% | 8.88% | 43.7% |
| US | 6.52% | 69.3% | 9.09% | 45.0% |

*Table K.2. Estimated human data savings from LLM-informed priors based on discrepancy between PRST and RST*

| | **Model with LLM-informed priors uses 2% of the human sample** | | **Model with LLM-informed priors uses 5% of the human sample** | |
|---|---|---|---|---|
| | Fraction of the human sample required for the model with uninformative priors | Sample size savings | Fraction of the human sample required for the model with uninformative priors | Sample size savings |
| China | 5.00% | 60.0% | 10.00% | 50.0% |
| Germany | 5.45% | 63.3% | 8.64% | 42.1% |
| India | 3.41% | 41.4% | 10.00% | 50.0% |
| Singapore | 4.22% | 52.6% | 20.00% | 75.0% |
| US | 8.00% | 75.0% | 8.00% | 37.5% |